\documentstyle[12pt,psfig]{article}
\newcommand{\resection}[1]{\setcounter{equation}{0}\section{#1}}
\newcommand{\appsection}{\addtocounter{section}{1} \setcounter{equation}{0}
                         \section*{Appendix \Alph{section}}}

\newcommand{\de}{\mbox{d}}
\newcommand{\be}{\begin{equation}}
\newcommand{\ee}{\end{equation}}
\newcommand{\EQ}{\begin{equation}}
\newcommand{\EN}{\end{equation}}
\newcommand{\th}{\theta}
\newcommand{\hs}{\hspace{1mm}}
\newcommand{\ov}{\overline}
\newcommand{\spz}{\hspace{3mm}}

\newcommand{\bd}{\begin{displaystyle}}
\newcommand{\ed}{\end{displaystyle}}
\newcommand{\ba}{\begin{array}}
\newcommand{\ea}{\end{array}}
\newcommand{\hb}{\overline{h}} 
\newcommand{\lb}{\overline{l}} 
\newcommand{\bb}{\overline{\beta}} 

\newcommand{\xb}{\overline{x}}
\newcommand{\yb}{\overline{y}}
\newcommand{\etab}{\overline{\eta}} 
\newcommand{\sigb}{\overline{\sigma}}
\newcommand{\F}{{\cal F}}
\newcommand{\st}{\stackrel}
\begin{document}
\setcounter{page}{0}
\topmargin 0pt
\oddsidemargin 5mm
\renewcommand{\thefootnote}{\arabic{footnote}}
\newpage
\setcounter{page}{0}

\begin{titlepage}

\begin{flushright}
ISAS/EP/95/161\\
\end{flushright}
\vspace*{0.5cm}

\begin{center}
{\bf
\begin{Large}
{\bf
Form Factors and Correlation Functions of the Stress--Energy
Tensor in Massive Deformation of the Minimal Models 
$\left( E_n \right)_1 \otimes\left( E_n \right)_1/\left( E_n \right)_2$
\\}
\end{Large}
}
\vspace*{1.5cm}{\large C. Acerbi$^{1,2}$,
         G. Mussardo$^{1,2,3}$
         and A. Valleriani$^{1,2}$}
         \\[.3cm]
         {\it $^1$International School for Advanced Studies}\\
          {\it $^2$Istituto Nazionale di Fisica Nucleare}\\
          {\it $^3$International Centre of Theoretical Physics }\\
          {\it 34013 Trieste, Italy}\\
          
\end{center}
\vspace*{0.7cm}

\begin{abstract}
\noindent
The magnetic deformation of the Ising Model, the thermal deformations 
of both the Tricritical Ising Model and the Tricritical Potts Model 
are governed by an algebraic structure based on the Dynkin diagram 
associated to the exceptional algebras $E_n$ (respectively for $n=8,7,6$). 
We make use of these underlying structures as well as of the 
discrete symmetries of the models to compute the matrix elements 
of the stress--energy tensor and its two--point correlation 
function by means of the spectral representation method. 
\end{abstract}
\vfill
\end{titlepage}
\setcounter{footnote}{0}
\resection{Introduction}

Important progress has recently been achieved in the computation of
correlation functions for integrable models,  defined either as  
lattice systems or as continuum theories. In addition to the well established
results on the spin--spin correlation function in the Ising model away from
 the critical temperature \cite{Ising1,McBook,Ising2,Ising3}, 
correlation functions of several important statistical integrable models have 
been obtained by means of different techniques, such as those 
discussed in the references 
\cite{Japan,Korepin}, for instance. Furthermore, for those 
models which present relativistic 
invariance and for which the exact $S$--matrix is known, a powerful method to 
compute the correlation functions is provided by the Form Factor (FF) approach 
originally proposed in \cite{KW,Smirnov}. This approach has proved to be 
extremely efficient because it leads to  fast convergent series for the 
correlators, as confirmed for instance in \cite{ZamYL, DM, DMS}. 
One of the most remarkable
results achieved by means of the FF approach is the solution of the 
long--standing problem of the computation of the spin--spin correlator 
of the Ising Model in a Magnetic Field at $T=T_c$ (IMMF) \cite{DM}. 
The aim of this paper is to extend the analysis of 
reference \cite{DM} to two statistical models which are very 
closely related to the 
IMMF, namely those relative to the thermal deformation of the Tricritical 
Ising model (TIM) and the Tricritical 3--state Potts model (TPM). 
The dynamics of all these systems are ruled by an 
algebraic structure related to the exceptional algebras $E_n$. In fact, 
the magnetic deformation of the Ising model 
highlights its underlying $E_8$ 
structure as well as the thermal deformation of the TIM and of 
the TPM, which highlights  
respectively the $E_7$ and the $E_6$ structures of these models. 
In this paper we are concerned with the determination of the FFs of one 
of the most important fields of the above mentioned models, i.e. 
the stress--energy tensor 
$T_{\mu\nu}(x)$. The matrix elements of this operator are particularly 
simple to determine for several reasons. Firstly, 
all the matrix elements of the components of this tensor can be 
expressed in terms of the FFs of just one scalar operator, namely its trace 
$\Theta(x)$. This operator, in turn, is related to the operator $\Phi(x)$ 
which deforms the conformal action by the relationship 
\begin{equation}
\label{tetafi}
\Theta(x) = 2\pi \lambda\, (2-2\Delta_{\Phi})\, \Phi(x)\,\, ,
\end{equation}
where $\Delta_{\Phi}$ is the conformal dimension of the field $\Phi(x)$. 
The stress--energy tensor is furthermore characterized by its conservation law 
$\partial^{\mu} T_{\mu\nu}(x)=0$ and by its obvious relation to the total 
energy and momentum of the system. These additional pieces of information 
are sufficient to uniquely identify the operator $\Theta(x)$ and then,  
to reconstruct its correlation functions\footnote{Identification of other 
local operators of an integrable theory may be in general more difficult. 
Although we are still lacking a general answer to this problem, nevertheless 
the series of papers \cite{JLG, KM} already yields some remarkable theoretical 
results.}. 
Some simple tests, based on certain sum rules, may be performed to check 
the efficiency of the  
approach. The first check involves the zeroth moment of the two--point 
correlation function $G(x) = \langle\,\Theta(x) \Theta(0)\,\rangle$, a quantity which is 
related to the amplitude $U$ of the free energy per unit 
volume $f_s\sim -U m^2$. Conversely, this amplitude can also be obtained
independently by means of the Thermodynamics Bethe Ansatz \cite{TBA,FateevTBA}. 
The second check is based on the second moment of $G(x)$ which is proportional 
to the difference of the central charges $c_{uv}$ and $c_{ir}$ of the 
conformal field theories arising in the ultraviolet and 
infrared limits \cite{cth}. 
An important point illustrated by these comparisons 
is that the spectral representation 
series of the correlation function is saturated with a high degree of 
accuracy by its first terms. This implies that 
the analytic efforts needed to 
obtain an accurate determination of the correlation function are simplified  
enormously\footnote{It is worth to notice that this remarkable behaviour 
of the spectral series has been also observed for massless models \cite{DMS}}. 

The paper is organized as follows: in the next section we briefly
review the bootstrap approach to integrable models based on the exact 
$S$--matrix formulation. In Section 3 and 4 we compute the first Form 
Factors of the stress--energy tensor of the perturbed Tricritical Ising 
Model and Tricritical Potts models respectively (the computations 
relative to the Ising model in a magnetic field may be found in the 
original reference \cite{DM}). Our conclusions are presented in Section 5. 
The paper also contains two appendices: in the first one we gather 
useful mathematical formulas used in our computations, while the second one 
contains the three--particle FF results of the TPM.

\resection{General Features of Integrable Quantum Field  \newline Theories}

In order to clarify the basic principles of the bootstrap approach to 
the solution of an integrable Quantum Field Theory and to set up the 
notation, in this section we endeavor to present a concise picture of 
the conceptual framework on which such approach is based and its most 
important consequences. Our starting point is the theory of integrable 
scattering processes, as originally developed in \cite{ZZ,Zam}.  

\subsection{Perturbed CFT and Factorized Scattering Theory}

The statistical systems analized in this article are particular 
integrable massive deformations of the first minimal unitary models of 
Conformal Field Theories (CFT), namely the Ising model, 
the Tricritical Ising Model and the Tricritical 3--state Potts Model. The 
Ising Model is deformed along the magnetic direction whereas the deformation 
of the other two models is along their thermal direction. The three models 
belong to the minimal unitary series of CFT and therefore, at their 
critical point they may be simply realized in terms of a coset contruction 
based on the affine $SU(2)$ algebra \cite{GKO}. However, it is well known 
that at criticality they also admit an alternative realization in terms of the
coset constructions 
$\left( E_n \right)_1 \otimes\left( E_n \right)_1/\left( E_n \right)_2$
based on the exceptional algebras $E_n$: the Ising Model is 
associated to the exceptional algebra $E_8$, the Tricritical Ising Model to 
$E_7$ and the Tricritical 3--state Potts Model to $E_6$. As shown in 
\cite{Zam,EY,FZ,ChMu,Sotkov}, the advantage of considering such alternative 
realizations of the three models at criticality becomes clear 
once they are perturbed away  from the fixed points along particular 
directions in the scaling region. Since we will have  the 
opportunity to come back 
and comment on this point in the next sections, where we consider in more 
detail each model, here we prefer to keep our discussion as general 
as possible and focus our attention on the common features of a typical 
integrable perturbed CFT. For such a theory, the action can be formally 
written as 
\be
{\cal A} \, =\, {\cal A}_{\rm CFT} + g\,\int d^2 x\, \Phi(x) \,\, ,
\label{pertaction}
\ee
where $\Phi(x)$ is the relevant operator that moves the system away  from 
criticality. It breaks the original conformal invariance and therefore 
introduces a mass scale in the theory. For integrable deformations, the 
dynamics of the system is supported by the presence of an infinite number 
of conserved charges (this may be argued by means of the counting argument 
proposed in \cite{Zam}). One of the consequences of the quantum integrability 
is that the scattering processes of the massive excitations of the theory 
are completely elastic. Since production processes are absent in the 
dynamics of such integrable QFT, their general $S$--matrix element is 
diagonal in the number of external ({\em in} and {\em out}) particles 
involved in the collision and moreover is completely factorized into the 
product of two--particle amplitudes. Hence, to know the 
on--mass--shell properties of such class of QFT we have simply to determine the 
two--particle $S$--matrices. For the class of models which we will consider 
in this paper, there is an additional simplification, i.e. all particles can be  
unambigously distinguished on the basis of their different quantum 
numbers (this is certainly the case for theories with a non degenerate 
mass spectrum). Under this circumstance, the scattering processes are 
completely diagonal and the two--particle scattering amplitudes are 
simply defined by the equation
\be
\mid A_a(\theta_a) \,A_b(\theta_b)\,\rangle_{\rm out} = S_{ab}(\theta_{ab}) 
\mid A_a(\theta_a) A_b(\theta_b)\,\rangle_{\rm in} \,\, ,
\ee
where $\theta_{ab} = \theta_a - \theta_b$ and we have 
used, as usual,  $\theta_a$ to parametrize the dispersion relation of the
particle $A_a$, i.e. 
\[
(E,p)= (m_a\cosh\theta_a,m_a\sinh\theta_a) \,\,\, .
\]
In terms of $\theta_{ab}$, the Mandelstam variable $s$ reads
$ s = m_a^2+m_b^2 +2 \, m_a m_b \cosh\theta_{ab}\,
$.
As functions of the variable $\theta_{ab}$ (from now on simply denoted 
by $\theta$), $S_{ab}(\theta)$ are analytic functions, with possible poles 
on the imaginary axis $0\leq\th\leq i\pi$. They satisfy the functional 
equations 
\be
\label{uncr}
\ba{c}
\bd
S_{ab}(\theta)\,S_{ab}(-\th)=1 \,\, ,
\ed
\\ 
\\
\bd
S_{ab}(i\pi-\th)=S_{\bar{a}b}(\th)\, ,
\ed
\ea
\ee
expressing the unitarity condition and the crossing symmetry of the theory,
respectively. The general solution of (\ref{uncr}) can be written in 
terms of an arbitrary product of the functions \cite{mitra}
\be\label{essealfa}
s_{\alpha}(\theta) = \frac{\sinh\frac{1}{2}(\theta+i\pi\alpha)}
{\sinh\frac{1}{2}(\theta-i\pi\alpha)}\, ,
\ee
where $-1\leq\alpha\leq 1$. The parameter $\alpha$ is related to the 
position of the pole of $s_{\alpha}(\theta)$ which is located at 
$\th=i\pi\alpha$. For those models with a non--degenerate mass spectrum 
(in which all particles are self--conjugate), the functional space of the 
solution of (\ref{uncr}) is instead spanned by an arbitrary product of 
the crossing symmetric functions
\be  
\label{effealfa}
f_{\alpha}(\theta)\, \equiv \,
s_{\alpha}(\th)s_{\alpha}(i\pi-\th) =
\frac{\tanh\frac{1}{2}(\theta+i\pi\alpha)}
{\tanh\frac{1}{2}(\theta-i\pi\alpha)}\,\,\,.
\ee
The simple poles of $f_{\alpha}$ are  located at the crossing 
symmetric points $\th=i\pi\alpha$ and $\th=i\pi (1-\alpha)$. 

For a theory with a degenerate mass spectrum the general expression of the 
two--particle S--matrix element may be written as
\begin{equation}
\label{smatdeg}
S_{ab}(\theta)=\prod_{\alpha\in {\cal A}_{ab}} s_{\alpha}(\theta)^{p_{\alpha}}
\,\, ,
\end{equation}
while for a theory with a non--degenerate mass spectrum we have 
\begin{equation}
\label{smat}
S_{ab}(\theta)=\prod_{\alpha\in {\cal A}_{ab}}
f_{\alpha}(\theta)^{p_{\alpha}}\,\,\,.
\end{equation}
In both cases, the exponents $p_{\alpha}$ denote the multiplicities of the 
corresponding poles identified by the indices $\alpha$. For those models which
present an underlying algebraic structure related to a Dynkin diagram, as for 
instance the three models investigated in this paper, the labels 
$\alpha\in{\cal A}_{ab}$ are integer multiples of $1/h$ where $h$ is the Coxeter 
number of the associated Lie algebra. The complete two--particle 
$S$--matrices of the Tricritical Ising and Potts Models in a thermal 
perturbation (originally computed in \cite{FZ,ChMu,Sotkov}) are reported for 
convenience in Tables (2) and (7) (the one relative to the IMMF may be found 
in the original reference \cite{Zam}). 

The two--particle elastic $S$--matrices considered in this paper 
present quite a rich pattern of poles in the complex $\theta$ plane. 
According to the analysis carried out by several groups 
\cite{ChMu,ColT,Braden,ChM}, the simple 
and higher odd--order poles are associated to the presence of bound 
states appearing as intermediate virtual particles in the scattering 
processes. Viceversa, all even--order poles are simply due to 
intermediate multi--scattering processes with no one--particle singularities.
For a simple pole of $S_{ab}(\theta)$ at $\theta= i u_{ab}^c$ 
corresponding to a bound state $A_c$, we can compute 
both the mass of the bound state by means of the equation 
\begin{equation}
m_{c}^2 = m_a^2+m_b^2 +2 m_a m_b \cos u_{ab}^{c} \,\, ,
\end{equation}
and the on--mass--shell three--point coupling constant $\Gamma_{ab}^c$
by taking the residue on the pole
\begin{equation}
\label{gamma}
-i \lim_{\theta\rightarrow iu_{ab}^{c}}
(\theta - iu_{ab}^{c})
S_{ab}(\theta)= (\Gamma_{ab}^{c})^2 \,\,\,.
\end{equation}
In the case of a double pole of the $S$--matrix placed at 
$\th= i \pi\alpha$, the associated residue is given by 
\begin{equation}
\label{doubp}
-i \lim_{\theta_{ab}\rightarrow i\varphi}(\theta_{ab}-i\varphi)^2
S_{ab}(\theta)= (\Gamma_{ad}^c\Gamma_{b\bar{d}}^e)^2
S_{ce}(i\gamma) \, ,
\end{equation}
where  $\gamma=\pi-u_{\bar{c}d}^{\bar{a}}-u_{\bar{e}\bar{d}}^{\bar{b}}$.  
It is beyond our scope to discuss here in more detail the multi--scattering
interpretation relative to the analytic structure of the $S$--matrix. 
The interested reader may consult the original literature quoted above 
for a complete account of the pole structure of the $S$--matrix. 
Some examples will be however provided in the next sections to 
enlighten the analytic structure of matrix elements of local operators.

\subsection{Correlation Functions and Form Factors}

One of the approaches that has proved to be extremely efficient 
in the computation of  
correlation functions for statistical models away  from criticality is the 
spectral representation method. For integrable models, this approach has 
been originally proposed in \cite{KW,Smirnov} and further analysed and 
applied by different groups [7-18]. The simplest example which illustrates this
approach is given by the computation of the two--point functions 
(higher--point functions being determined similarly). In the spectral 
representation approach, the two--point function of a local operator 
$\varphi(x)$ may be expressed as   
\begin{equation}
\label{formexp}
\langle\,\varphi(x)\varphi(0)\,\rangle =
\sum_{n=0}^{\infty}\int_{\th_1>\th_2\ldots>\th_n}\frac{\de\theta_1\cdots\de\theta_n}
{(2\pi)^n}\left|F_{a_1,\ldots,a_n}^{\varphi}
(\theta_1,\ldots,\theta_n)\right|^2
e^{-|x|\sum_{k=1}^{n}m_k\cosh\theta_k} \,\,\, ,
\ee
where 
\be
\label{FF}
F_{a_1,\ldots,a_n}^{\varphi}(\theta_1,\ldots,\theta_n) \,\equiv\,
\langle\,0\,|\,\varphi(0)\,\mid
A^{\dagger}_{a_1}(\theta_1)\cdots A^{\dagger}_{a_n}(\theta_n)\,\rangle\,\,\, .
\ee
The above matrix elements are the so--called Form Factors (FF) and, as we 
will briefly discuss below, their computation can be performed once the 
exact $S$--matrix and the bound state structure of the theory are known. 
Inserted into (\ref{formexp}), they give rise to fast convergent series 
both in the infrared region ( $| x | \rightarrow \infty$), with 
a corresponding exponential decay, and in the ultraviolet limit 
($| x | \rightarrow 0$) where the correlation 
functions show power--law behaviours. 

The FF satisfy the so--called Watson equations, given by 
\begin{eqnarray} 
F^{\varphi}_{a_1,\ldots,a_i,a_{i+1},\ldots,a_n}(\theta_1,\ldots,
\th_i,\th_{i+1},\ldots,\th_n)\!\!\!&=&\!\!\!   \nonumber \\
  & &\!\!\!\!\!\!\!\!\!\!\!\!\!\!\!\!\!\!\!\!\!
  \!\!\!\!\!\!\!\!\!\!\!\!\!\!\!\!\!\!\!\!\!
   = S_{a_i,a_{i+1}}(\th_i-\th_{i+1})
\:F^{\varphi}_{a_1,\ldots,a_{i+1},a_i,\ldots,a_n}(\theta_1,\ldots,
\th_{i+1},\th_{i},\ldots,\th_n)\, ,\nonumber
\end{eqnarray}
\EQ \label{sf}
F^{\varphi}_{a_1,a_2,\ldots,a_n}(\theta_1+2\pi i,\th_2,\ldots,\th_n)= 
F^{\varphi}_{a_2,\ldots,a_n,a_1}(\th_2, \ldots, \th_n, \theta_1)\, .
\EN
Among the solutions of these equations, there are those (called {\em minimal 
solutions}) characterized by the property that they have neither poles nor 
zeros in the strips $ \mbox{\it Im}\, \th_{ij} \in (0,2\pi)$. 
By using the factorization properties of the underlying scattering 
theory, the minimal solution associated to a generic FF may be easily
expressed in terms of the minimal two--particle FFs 
$F_{ab}^{min}(\th)$ by 
\be
F_{a_1,a_2,\ldots,a_n}^{min}(\theta_1,\th_2,\ldots,\th_n) =
\prod_{1\leq i < j\leq n}F_{a_i a_j}^{min}(\th_i-\th_j) \,\,\,.
\ee
where the explicit expressions of $F_{ab}^{min}(\th)$ are given 
for theories with degenerate mass spectrum by  
\be
\label{fminE6}
F_{ab}^{min}(\th)=\left(-i\sinh\frac{\th}{2}\right)^{\delta_{a,b}}
\prod_{\alpha\in {\cal A}_{ab}} h_{\alpha}(\theta)^{p_{\alpha}} \,\, ,
\ee
while for theories with non--degenerate mass spectrum by 
\be
\label{fmin}
F_{ab}^{min}(\th)=\left(-i\sinh\frac{\th}{2}\right)^{\delta_{a,b}}
\prod_{\alpha\in {\cal A}_{ab}} g_{\alpha}(\theta)^{p_{\alpha}} \,\,\, .
\ee
The definition and the properties of the functions $h_{\alpha}(\theta)$ 
and $g_{\alpha}(\theta)$ are collected in Appendix A. 

The minimal expression of the FFs does not carry any dependence on the 
specific operator we are considering, as it must be, since the monodromy 
properties derive  from the $S$--matrix alone. To characterize the different 
operators and to take into account the dynamical pattern of bound states of 
the theory, let us consider in more detail the analytic structure 
of the FFs, starting our discussion from the occurrence of their poles. 
Their pattern may be very complicated for the multi--scattering 
processes of the theory. There are however two classes of simple order 
poles in the FF which have a simple and natural origin \cite{Smirnov}. The 
first class is that of kinematical poles relative to particle--antiparticle 
singularities at the relative rapidity $\theta = i \pi$ with the corresponding 
residue given by  
\be 
\label{kinpole}
-i\lim_{\tilde{\th}\rightarrow\th}F_{\bar{a},a,a_1,\ldots,a_n}
(\tilde{\th}+ i\pi,\th,\th_1,\ldots,\th_n)=\left(1-\prod_1^n S_{aa_i}(\th-\th_i)\right)
F_{a_1,\ldots,a_n}(\th_1,\ldots,\th_n)\,\,\,.
\ee
The second class of simple order poles of the FFs which admit a simple 
explanation is that associated to bound state singularities. 
Namely, whenever $A_a(\theta_a)$ and $A_b(\theta_b)$ form a bound state 
$A_c(\theta)$ for the value $\theta_{ab} = i u_{ab}^c$ of their relative 
rapidity, then all the matrix elements
$F^{\varphi}_{a,b,a_1\ldots,a_n}(\theta_a,\th_b,\th_1,\ldots,\th_n)$
involving the two particles $A_a(\theta_a)$ and $A_b(\theta_b)$ 
will have as well a simple order pole at the same position, 
with the residue ruled by the on--shell three--point coupling 
constant $\Gamma_{ab}^c$,  i.e. (see Figure 1)
\be
\label{boundfpolebis}
-i\lim_{\th_{ab}\rightarrow iu_{ab}^{c}}(\th_{ab}-iu_{ab}^{c})\,
F^{\varphi}_{a,b,a_1,\ldots,a_n}(\th_a,\th_b,\th_1,\ldots,\th_n)
=\,\Gamma_{ab}^{c}\,
F^{\varphi}_{c,a_1,\ldots,a_n}(\th_c,\th_1,\ldots,\th_n)\,\,\,.
\ee
In addition to the above classes of simple poles, the FFs may present 
poles of higher order relative to the underlying multi--scattering 
processes, as recently clarified in \cite{DM}.  

A key point to understand the rich analytic structure of the matrix 
elements is to initially\footnote{
The reason is that, by factorization, FFs with higher number of 
particles inherit their pole structure  from the analytic structure of the 
two--particle channels. Moreover, the two--particle FFs play an 
important role in the theory since they provide the ``initial conditions'' 
needed for solving the recursive functional equations (\ref{kinpole}) 
and (\ref{boundfpolebis}).} analyse the two--particle FFs. 
Following the analysis of \cite{DM} (see also \cite{Oota}), the two--particle 
FFs can be conveniently written as
\be
\label{Fphi}
F_{ab}^{\varphi}(\th) = Q_{ab}^{\varphi}(\th)
\frac{F_{ab}^{min}(\th)}{D_{ab}(\th)} \,\, ,
\ee
where $D_{ab}$ takes into account its poles structure and $Q_{ab}^{\varphi}$ 
is a polynomial in $\cosh\th$ which carries the dependence on the operator 
$\varphi$.

The polynomials $D_{ab}(\th)$ are determined  from the poles of the 
$S$--matrix. The analysis of ref.\,\cite{DM} gives the following simple rules 
for determining them in the case of non--degenerate  theories: 
\be
D_{ab}(\th)=\prod_{\alpha\in {\cal
A}_{ab}} \Bigl({\cal P}_\alpha(\th)\Bigr)^{i_\alpha}
\Bigl({\cal P}_{1-\alpha}(\th)\Bigr)^{j_\alpha} \,\,\,,
\label{dab}
\ee 
\be
\begin{array}{lll}
i_{\alpha} = n+1\,\,\, , & j_{\alpha} = n \,\,\, , &
\mbox{\rm if} \hspace{1cm} p_\alpha=2n+1\,\,\,; \\
i_{\alpha} = n \,\,\, , & j_{\alpha} = n \,\,\, , &
\mbox{\rm if} \hspace{1cm} p_\alpha=2n\,\,\, ,
\end{array}
\ee
where ${\cal A}_{ab}$ and $p_\alpha$ are defined in eq. 
(\ref{smat}). The functions
\be
\label{pmin}
{\cal P}_{\alpha}(\th) \equiv \frac{\cos\pi\alpha - \cosh\th}
{2 \cos^2\frac{\pi\alpha}{2}}\, 
\ee
give a suitable parametrization of the pole at $\th=i\pi\alpha$.
The above prescription can be also generalized to degenerate
theories. In fact, referring to equation (\ref{essealfa}), one can write
\be
D_{ab}(\th)=\prod_{\alpha\in {\cal
A}_{ab}} \Bigl({\cal P}_\alpha(\th)\Bigr)^{i_\alpha}
\,\,\,,
\label{dabE6}
\ee
\be
\begin{array}{lll}
i_{\alpha} = n+1\,\,\, ,  &
\mbox{\rm if} \hspace{1cm} p_\alpha=2n+1\,\,\, & \mbox{\rm $s$--channel pole};
\\
i_{\alpha} = n\,\,\, ,  &
\mbox{\rm if} \hspace{1cm} p_\alpha=2n+1\,\,\, & \mbox{\rm $t$--channel pole}; 
\\
i_{\alpha} = n \,\,\, ,  &
\mbox{\rm if} \hspace{1cm} p_\alpha=2n\,\,\, .
\end{array}
\ee
where it is convenient to distinguish between poles associated to the direct 
$s$--channel and those relative to the crossed $t$--channel. As we will show 
in the sequel, the above rules play an essential role for implementing 
the boostrap program for the computation of the FFs in the TIM and 
in the TPM. Let us quote at this point the equations which will be often 
employed in the next sections. Those are: (a) the residue equations 
at a simple order pole that corresponds to a bound state  
\be
\label{boundfpole}
-i\lim_{\th \rightarrow iu_{ab}^{c}}(\th -iu_{ab}^{c})
F^{\varphi}_{ab}(\th)=
\Gamma_{ab}^{c}
F^{\varphi}_{c} \,\, ,
\ee
(see Figure 1); (b) the residue equations relative to a simple order pole 
induced by a double pole in the $S$--matrix 
\be
\label{doubfpole}
 - i \lim_{\th_{ab}\rightarrow i\varphi}(\th_{ab}-i\varphi) 
F_{ab}(\th_{ab}) \,=\,
\Gamma_{ad}^c\Gamma_{\bar{d}b}^e F_{ce}(i\gamma)\,\, ,
\ee
where $\gamma=\pi- u_{\bar{c}d}^{\bar{a}}-u_{\bar{d}\bar{e}}^{\bar{b}}$ 
(see Figure 2) and finally, (c)  
the residue equations relative to a double order pole induced by 
a third order pole in the corresponding $S$--matrix (see Figure 3 where $\varphi=u_{ab}^{f}$)
\be
\label{tripfpole}
\lim_{\th_{ab}\rightarrow i u_{ab}^{f}}
(\th_{ab}-i u_{ab}^{f})^2 
F_{ab}(\th_{ab}) =
 i \Gamma_{ad}^c\Gamma_{b\bar{d}}^e
\lim_{\th_{ce}\rightarrow i u_{ce}^{f}}
(\th_{ce}-i u_{ce}^{f})
F_{ce} (\th_{ce}) =  - \Gamma_{ad}^c\Gamma_{b\bar{d}}^e\Gamma_{ce}^{f} F_{f} \,\, .
\ee 
After having considered the pole structure of the two--particle FFs, 
let us concentrate our attention on the 
polynomial $Q_{ab}^{\varphi}(\th)$ in the numerator of (\ref{Fphi}). 
In contrast to $D_{ab}(\th)$, which is only fixed by the $S$--matrix 
singularities, 
the polynomials $Q_{ab}^{\varphi}(\th)$ depend, on the contrary, 
on the operator $\varphi(x)$ and may be used to characterize it. An 
upper bound on the maximal degree of the polynomials $Q_{ab}^{\varphi}(\th)$ 
has been derived in \cite{DM}. Briefly stated, the argument consists in 
looking at the large energy limit of the FF and relating it to the conformal 
properties of the corresponding operator $\varphi(x)$. Denoted by 
$\Delta_{\varphi}$ its conformal weight and 
by $y_{\varphi}$ the real quantity defined by 
\[
\lim_{\mid \th_i\mid \rightarrow \infty} F^{\varphi}
_{a_1,\ldots,a_n}(\th_1,\ldots,\th_n)\,\sim e^{y_{\varphi}\mid\theta_i\mid}
\]
we have \cite{DM}
\EQ 
\label{boundff}
y_{\varphi} \leq \Delta_{\varphi} \,\,\,.
\EN
Taking into account the degree of the factor $F_{ab}^{min}/D_{ab}(\theta)$ in 
the two--particle FF (\ref{Fphi}) by also using eq. (\ref{gas}),
it is easy to translate the 
inequality (\ref{boundff}) into an upper bound on the degree of the polynomial 
$Q_{ab}^{\varphi}(\theta)$. 

In what follows we will consider the FFs of the trace 
$\Theta(x)$ of the stress--energy tensor. In this case we have 
additional constraints 
for the corresponding polynomial $Q_{ab}(\th)$. 
In fact, the conservation law $\partial^{\mu} T_{\mu \nu}(x)$ satisfied 
by the stress--energy tensor
implies that the FFs of the trace $\Theta(x)$ must contain the kinematical
polynomial $P^2=(p_1+\cdots+p_n)^2$, with the exception of the FFs with 
two identical particles. For the two--particles FFs, this property 
can be expressed 
by means of the following factorization
\be
\label{qpol}
Q^{\Theta}_{ab}(\th)\,=\,
\left(\cosh\th+\frac{m_a^2+m_b^2}{2m_a m_b}\right)^{1 - \delta_{ab}}
\, P_{ab}(\th) \,\, ,
\ee
where
\be
\label{ppol}
P_{ab}(\th) \,\equiv\,\sum_{k=0}^{N_{ab}}a_{ab}^k\cosh^k\th\,\, ,
\ee
The degree $N_{ab}$ in (\ref{ppol}) may be determined by implementing 
the inequality 
(\ref{boundff}). In this way, the problem is reduced to determine 
the coefficients $a_{ab}^k$ of the polynomials $P_{ab}$. This goal 
can be achieved 
by applying the residue equations together with 
the normalization condition of the two--particle FF, which are expressed by 
\be
\label{norm}
F^{\Theta}_{a\bar{a}}(i\pi)= 2\pi m_a^2\, .
\ee
The above conditions prove in general sufficient or even redundant in number, 
to fix all the coefficients of the polynomial $P_{ab}(\th)$.  

As mentioned in the introduction, there is now a strong evidence that 
the spectral series based on the FFs are fastly convergent. 
For the correlation functions of the stress--energy tensor, one way to 
test this convergence is to employ two sum rules satisfied by the moments 
${\cal M}_p$ of the two--point function of $\Theta(x)$, defined by 
\EQ
{\cal M}_p \,=\,\int d^2 x \,|x|^p \,\langle\,\Theta(x) \Theta(0)\,\rangle \,\,\,.
\EN 
The first sum rule is relative to the bulk free energy $f\sim -U m^2$,  
where the amplitude $U$ is related to the zero--moment ${\cal M}_0$ by 
\be
\label{usum}
U \,=\, \frac{1}{16 \Delta_{\Phi}}\,\frac{1}{\pi^2 m^2}\,{\cal M}_0\,\, ,
\ee
with $m$ the lightest mass of the theory. The second sum rule relates the 
second moment ${\cal M}_2$ to the central charge $c$ of the original CFT, 
according to the formula 
\cite{cth}
\be
\label{csum}
c\,=\, \frac{3}{4\pi} {\cal M}_2 \,\,\,.
\ee
By inserting the spectral representation of the two--point function 
$\langle\,\Theta(x) \Theta(0)\,\rangle$, both moments can be expressed as a series on the 
number of the intermediate particles with an increasing value of their 
center--of--mass energy. In this way we can test the fast convergence of 
the spectral representation by comparing the truncated values of 
the series of the moments ${\cal M}_0$ 
and ${\cal M}_2$, with  the known value of the central charge $c$ and 
the exact value of $U$ computed by means of the Thermodinamic 
Bethe Ansatz \cite{TBA,FateevTBA}.

\resection{Form Factors of the Energy Operator for the \newline Thermal~Perturbation 
of the  Tricritical \newline Ising Model}

The Tricritical Ising model is the second model in the minimal unitary
conformal series with central charge $c=7/10$ and four relevant 
fields \cite{BPZ}. The microscopic formulation of the model, its 
conformal properties 
and its scaling region nearby the critical point have been discussed in 
several papers (see, for instance 
\cite{FZ,ChMu,Blume,Lawrie,vongehlen,Shen1}). 
In the following we give a short review of the features of the TIM
which are most relevant to the FF approach to integrable 
massive models.

\subsection{Generalities of the TIM} 

The Tricritial Ising model  may be regarded as the 
universality class of the Landau--Ginzburg $\Phi^6$--theory 
\EQ
L \,=\,(\nabla \Phi)^2 + g_6 \Phi^6 + g_4 \Phi^4 + g_3 \Phi^3 + g_2 \Phi^2+
g_1 \Phi
\label{Landau}
\EN
at its critical point $g_1=g_2=g_3=g_4=0$ \cite{ZamLG}. This Lagrangian 
describes the continuum limit of microscopic models with a tricritical point, 
among them the Ising model with annealed vacancies, with an Hamiltonian given 
by \cite{Blume,Lawrie} 
\EQ 
{\cal H}=-\beta \sum_{<ij>}\sigma_i\sigma_j t_i t_j-\mu\sum_i t_i 
\hs\hs.
\EN
$\beta$ is the inverse temperature, $\mu$ the chemical potential, $\sigma_i = 
\pm 1$ the Ising spins and  $t_i = 0,1$ is the vacancy variable. The model 
has a tricritical point $(\beta_0,\mu_0)$ related to the spontaneous 
symmetry breaking of the $Z_2$ symmetry. At the critical point 
$(\beta_0,\mu_0)$, the TIM can be described by the following scaling 
fields: the energy density $\epsilon(z,\overline{z})$ with anomalous 
dimensions $(\Delta,\overline{\Delta})=(\frac{1}{10},\frac{1}{10})$, the 
vacancy operator or subleading energy operator $t(z,\overline{z})$ 
with $(\Delta,\overline{\Delta})=(\frac{3}{5},\frac{3}{5})$,
the irrelevant field $\epsilon^{\prime\prime}$ with $(\Delta,\overline{\Delta})=
(\frac{3}{2},\frac{3}{2})$, the magnetization field (or order--parameter) 
$\sigma(z,\overline{z})$ with $(\Delta,\overline{\Delta})=
(\frac{3}{80},\frac{3}{80})$, and the so--called subleading magnetization 
operator $\alpha(z,\overline{z})$ with anomalous dimensions 
$(\frac{7}{16},\frac{7}{16})$. With rescpect  to the $Z_2$ symmetry of the spin 
model, the spin operators are odd while the energy operator, the vacancy 
operator and the irrelevant field $\epsilon^{\prime\prime}$ are even. 

A peculiar feature of the TIM is the presence of another infinite 
dimensional symmetry in addition to the Virasoro algebra, i.e. 
a hidden $W$-algebra based on the $E_7$ root system. This is related to the 
equivalent construction of the TIM in terms of the coset model 
$(\hat E_7)_1\otimes(\hat E_7)_1/(\hat E_7)_2$. Let us briefly recall the 
coset formulation at the critical point \cite{GKO}.  From the 
theory of  Kac--Moody algebras, the central charge of a CFT 
constructed on an affine Lie algebra $G$ at level $k$ is given by
\EQ
c_{G}=\frac{k\mid G\mid}{k + h_G} \hs\hs, 
\EN
where $\mid G\mid $ is the dimension of the algebra and $h_G$ the
dual Coxeter number. The unitarity condition for the CFT restricts the highest
weight representations $\mid \lambda\,\rangle$ which can appear at the level $k$. 
Denoting with $\omega$ the highest root, the allowed representations 
$\mid\lambda \,\rangle$ at the level $k$ must satisfy
\EQ
\frac{2\,\omega\cdot\lambda}{\omega^2}\leq k \hs\hs, 
\EN
and their dimension is given by 
\EQ
\Delta_{\lambda}=\frac{C_{\lambda}/\omega^2}{k + h_G} \hs\hs, 
\label{dimens}
\EN
where $C_{\lambda}$ is the quadratic Casimir in the representation 
$\{\lambda\}$. Using a subgroup $H \subset G$, one can construct a CFT 
on the coset group $G/H$, with a central charge equal to 
\EQ
c_{G/H}=c_{G}-c_{H} = {{k_G \mid G\mid}\over{k_G + h_G}}
-{{k_H\mid H\mid}\over{k_H + h_H}} \,\,\,.
\label{cccentralcharge}
\EN
Its representations $\psi^k$ are simply obtained by the decomposition of the 
Hilbert space
\EQ
\mid c_G,\lambda_G \,\rangle = \spz \oplus_k\spz \left[
\spz \mid c_{G/H}, \psi^k_{G/H} \,\rangle
\otimes \mid c_{H}, \lambda_H^k \,\rangle\spz\right] \hs\hs. 
\label{decomp}
\EN
In the case of the TIM, $h = 18$ and eq.\hs(\ref{cccentralcharge}) 
gives $c=\frac{7}{10}$. At level $k=1$, the possible representations are the 
identity $1$ and the representation $\Pi_6$ with scaling dimension $0$ and 
$\frac{3}{4}$ respectively
\EQ
\begin{array}{ccl}
(E_7)_1 &\rightarrow &\hspace{3mm} \{1,\Pi_6\}=\{0,\frac{3}{4}\} \,\,\,.
\end{array}
\EN
Their components $(n_1,n_2,\cdots,n_7)$ ($n_i$ integer) with respect to the
simple roots of $E_7$ are given by \cite{Slansky}
\EQ
\begin{array}{ccl}
1 &\rightarrow& (0,0,0,0,0,0,0)\\
\Pi_6 &\rightarrow& (0,0,0,0,0,1,0) \hs\hs. 
\end{array}
\EN
At the level $k=2$, the representations are given by 
\EQ
\begin{array}{ccl}
(E_7)_2 &\rightarrow &\hspace{3mm} \{1,\Pi_1,\Pi_2,\Pi_5,\Pi_6\}=
\{0,\frac{9}{10},\frac{21}{16},\frac{7}{5},\frac{57}{80}\}
\end{array} \hs\hs ,
\EN
with the corresponding fundamental weights 
\EQ
\begin{array}{ccl}
\Pi_1 &\rightarrow & (1,0,0,0,0,0,0) \,\,\, ,\\
\Pi_2 &\rightarrow & (0,1,0,0,0,0,0) \,\,\, ,\\
\Pi_5 &\rightarrow & (0,0,0,0,1,0,0)\,\,\,. 
\end{array}
\EN
$\Pi_1$ is the adjoint representation. Using eq.\,(\ref{decomp}), the scaling 
dimensions of the TIM are recovered by the decomposition 
\EQ
\begin{array}{ccl}
(0)_1\times (0)_1 &=& [(0)_{TIM} \,\otimes \,(0)_2] \, +
\,[(\frac{1}{10})_{TIM} \otimes (\Pi_1)_2] \, + \,
[(\frac{6}{10})_{TIM} \otimes (\Pi_5)_2] \,\, ,\\
(0)_1 \times (\frac{3}{4})_1 &=& [(\frac{7}{16})_{TIM}
\otimes (\Pi_2)_2]\, + \, [(\frac{3}{80})_{TIM} \otimes (\Pi_6)_2] \,\, ,\\
(\frac{3}{4})_1 \times (\frac{3}{4})_1 &=&
(\frac{3}{2})_{TIM} \otimes (0)_2 \,\,\,.
\end{array}
\label{tensor}
\EN

The off--critical perturbation considered in this paper is the one 
given by the leading energy operator $\epsilon(z,\overline z)$ of conformal 
weights  $\left(\frac{1}{10},\frac{1}{10}\right)$. Note that this operator 
is associated to the adjoint of $E_7$. According to the analysis of 
\cite{EY}, this leads to a structure of the off--critical system 
deeply related to the root system of $E_7$, as we briefly recall in the 
following.  

First of all, the off--critical massive model shares the same grading of 
conserved currents as the Affine Toda Field Theory constructed on the root 
system of $E_7$, {\em i.e.} the spins of the higher conserved currents are 
equal to the exponents of the $E_7$ algebra modulo its Coxeter number 
$h=18$, i.e.
\EQ
s=1,5,7,9,11,13,17 \hspace{3mm} (\mbox{mod} \hs 18) \,\,\,.
\EN
The presence of these higher conserved currents implies the elasticity of the 
scattering processes of the massive excitations. To compute the mass spectrum 
and the scattering amplitudes, it is important to observe that, according to 
the sign of the coupling constant $g$ in (\ref{pertaction}), this 
perturbation drives the system either in its high--temperature phase or in its 
low--temperature phase. While in the latter phase we have a spontaneously 
symmetry breaking of the $Z_2$ symmetry of the underlying microscopic spin 
system, in the former phase the $Z_2$ symmetry is a good quantum number and 
therefore can be used to label the states. In the low--temperature phase,
the massive excitations are given by kink states and bound state thereof, in 
the high--temperature phase we have instead ordinary particle excitations. 
The two phases are related by a duality transformation and therefore we can 
restrict our attention to only one of them, which we choose to be the 
high--temperature phase. In this phase, the massive excitations are given by 
seven self--conjugated particles $A_1,\ldots,A_7$ with mass 
\begin{eqnarray}
m_1 &=& M(g)\,, \nonumber \\
m_2 &=& 2\, m_1 \cos\frac{5\pi}{18} = (1.28557..) \,m_1\,,\nonumber\\
m_3 &=& 2\, m_1 \cos\frac{\pi}{9} = (1.87938..) \,m_1\,,\nonumber\\
m_4 &=& 2\, m_1 \cos\frac{\pi}{18} = (1.96961..) \,m_1\,,\nonumber \\
m_5 &=& 2\, m_2 \cos\frac{\pi}{18} = (2.53208..) \,m_1\,,\\
m_6 &=& 2\, m_3 \cos\frac{2\pi}{9} = (2.87938..) \,m_1\,,\nonumber\\
m_7 &=& 4\, m_3 \cos\frac{\pi}{18} = (3.70166..) \,m_1\,\,\,.
\nonumber
\end{eqnarray}
The dependence of the mass scale $M$ on the coupling constant $g$ has 
been computed in \cite{FateevTBA}
\EQ
M(g) \,=\, {\cal{C}} \, g^{\frac{5}{9}} ,
\EN
where
\EQ
 {\cal{C}} \,=\, \left[ 
   4 \, \pi^2 \:  \gamma(\mbox{$\frac{4}{5}$}) \,\gamma(\mbox{$\frac{3}{5}$})\,\gamma^2(\mbox{$\frac{7}{10}$}) 
               \right]^{\frac{5}{18}}
                \frac{2 \,\Gamma(\frac{2}{9})\,\Gamma(\frac{19}{18}) }
                {\Gamma(\frac{1}{2})\,\Gamma(\frac{2}{3})\,\Gamma(\frac{10}{9})} = 3.745372836\ldots,
\EN
where $
\gamma(x) \equiv \frac{\Gamma(x)}{\Gamma(1-x)} \,.
$
The mass ratios are proportional to the components of the Perron--Frobenius 
eigenvector of the Cartan matrix of the exceptional algebra $E_7$ \cite{Braden} 
 and therefore the particles $A_i$ may be put in correspondence 
with the following representations of $E_7$ (here identified by 
their dimensions) 
\EQ
\begin{array}{ccl}
A_1 &\rightarrow & 56 \hs\hs,\\
A_2 &\rightarrow & 133\hs\hs,\\
A_3 &\rightarrow & 912\hs\hs,\\
A_4 &\rightarrow & 1539\hs\hs,\\
A_5 &\rightarrow & 8645\hs\hs,\\
A_6 &\rightarrow & 27664\hs\hs,\\
A_7 &\rightarrow & 365750\hs\hs. 
\end{array}
\EN
The exact $S$--matrix of the model is given by the minimal $S$--matrix of the 
Affine Toda Field Theory based on the root system of $E_7$. It has been 
calculated in \cite{FZ,ChMu} and is listed for convenience in 
Table 2. The structure of the bound states 
may be written in a concise way by grouping the particle states 
into two triplets and one singlet states \cite{ChMu}
\EQ
\begin{array}{rcl}
(Q_1, Q_2, Q_3)\:\: &\equiv& \:\:(A_6, A_3, A_1) \,\,, \\
(K_1, K_2, K_3)\:\: &\equiv&\:\: (A_2, A_4, A_7)\,\,,  \\
(N) \:\: &\equiv& \:\:(A_5) \hs\hs. 
\end{array}
\EN
The first triplet consists of the $Z_2$ odd particles whereas the other triplet 
and the singlet are made of $Z_2$ even particles. The ``bootstrap fusions'' 
involving $[N]$ and $[N,K_i]$ form closed subsets 
\EQ
\begin{array}{lllllll}
N \cdot N &=& N \,, & &
N \cdot K_A &=& K_1 + K_2 + K_3\,, \\
K_A \cdot K_{A+1} &=& K_A + N\,,  & &
K_A \cdot K_A &=& K_A + K_{A+1} + N  \,\,.
\end{array}
\label{structure1}
\EN
Including the  first triplet, we obtain the following algebra
\EQ
\begin{array}{lllllll}
K_A \cdot Q_A &=& Q_{A+1}\,, & &
K_A \cdot Q_{A+1} &=& Q_1 + Q_2 + Q_3\,, \\
K_A \cdot Q_{A-1} &=& Q_{A-1} + Q_{A+1}\,, & &
Q_A \cdot Q_A &=& K_{A-1} + K_{A+1}\,, \\
Q_A \cdot Q_{A+1} &=& K_A + K_{A-1} + N\,, & &
N \cdot Q_A &=& Q_{A-1} + Q_{A+1} \,.
\end{array}
\label{structure2}
\EN
It has been observed that 
these bootstrap fusions are a subset of the tensor 
product decomposition of the associate representations of $E_7$ \cite{Braden}.

\subsection{Form Factors of the TIM}
After the discussion on the general features of the model, let us consider 
now the problem of computing the FFs of the operator $\epsilon(x)$ 
or, equivalently, of the trace $\Theta(x)$ of the stress--energy tensor. To 
this aim, the $Z_2$ parity of the model is extremely helpful. In fact, because of 
the even parity of the energy operator, we can immediatly conclude that its FF 
with a $Z_2$--odd (multi--particle) state must vanish. In particular, 
the one--particle FFs of $\Theta$ for the odd particles are all zero.

To start with the bootstrap procedure, let us consider the two--particle 
FF relative to the fundamental excitation $A_1$ 
\be
\label{f11e7}
F_{11}^{\Theta}(\th)=
\frac{F_{11}^{min}(\th)}{D_{11}(\th)}\,\,Q_{11}^{\Theta}(\th) \,\, ,
\ee
where
\be
\label{f11min7}
F_{11}^{min}(\th)=-i\sinh(\th/2)\,\,
g_{5/9}(\th)\,\,g_{1/9}(\th) \,\, ,
\ee 
and
\be
\label{D11min7}
D_{11}(\th)= {\cal P}_{5/9}(\th)\,\, {\cal P}_{1/9}(\th)\, .
\ee
By using the bound (\ref{boundff}), we see that the polynomial 
$Q_{11}^{\Theta}(\th)$ reduces just to a constant,  
which can be easily determined by means of the normalization 
condition (\ref{norm}), i.e. $a_{11}^0=2\pi m_1^2$. Thus $F_{11}(\th)$ 
is now completely determined and its expression can be used 
to derive the one--particle
FFs $F_2$ and $F_4$. Indeed, the particles $A_2$ and $A_4$ appear as
bound state of the particle $A_1$ with itself, the coupling $\Gamma_{11}^2$
and $\Gamma_{11}^4$ being easily determined by the residue equation 
(\ref{gamma}). By using then the equation for the bound state poles of the Form
Factors (\ref{boundfpole}), one gets the desired result (see Table 4).

To proceed further, it is convenient to list the $Z_2$ even states 
(the only ones giving non--vanishing FFs of the stress--energy tensor) in order 
of increasing energy, as in Table 3. 
After computing $F_{22}^{\Theta}$ , $F_{5}^{\Theta}$  and $F_{13}^{\Theta}$, 
which are obtained by means of the same technique, (i.e. fixing the unknown 
coefficients of FFs by using the simple pole residue equations), a more 
interesting computation is represented by the two--particle FF $F_{24}(\th)$. 
The corresponding $S$--matrix  element displays 
a double pole and therefore, according to eq. (\ref{dab}), we 
have 
\be
F_{24}^{\Theta}(\th)=\frac{F_{24}^{min}(\th)}
{D_{24}(\th)}\,\,Q_{24}^{\Theta}(\th)\,\, ,
\ee
where
\be
F_{24}^{min}(\th) = g_{7/9}(\th)\,\,g_{4/9}(\th)\,\,g_{1/3}^2(\th) \,\,\, ,
\ee
and
\be
D_{24}(\th) =  {\cal P}_{7/9}(\th)\,\, 
{\cal P}_{4/9}(\th)\,\, {\cal P}_{1/3}(\th)\,\, {\cal P}_{2/3}(\th)\,\,\,.
\ee
Taking into account the asymptotic behaviour of the FF and 
eqs. (\ref{qpol}) and (\ref{ppol}), we conclude that in this case
the polynomial 
$P_{24}$ has degree $N_{24}=1$ and therefore $Q_{24}(\th)$ reads 
\be
Q_{24}^{\Theta}(\th) \,=\, \left(\cosh\th +\frac{m_2^2+m_4^2}{2m_2 m_4}\right)
(a_{24}^0+a_{24}^1\cosh\th) \, .
\ee
To determine the constants $a_{24}^0$ and $a_{24}^1$, we need at least two 
linearly independent equations, which are provided by eq. (\ref{boundfpole})
on the fusions 
\be
\ba{ccc}
(A_2\, ,\, A_4)\rightarrow A_2 &
\mbox{ and } &
(A_2\, ,\, A_4)\rightarrow A_5\, .
\ea
\ee
Both $F_2$ and $F_5$ are known, of course, from previous computations.
In this case, the double pole in the $S$-matrix provides a non--trivial 
check for the computation. In fact, we have the process drawn in Figure 2, 
with the identification 
\[
a=2, \hspace{1cm} b=4, \hspace{1cm} d=e=1\,\, ,
\]
and respectively
\[ 
 c=1, \hspace{1cm} \varphi= 2\pi/3, \hspace{1cm} \gamma= \pi/3 \,\, ,
\] 
or
\[
c=3, \hspace{1cm} \varphi= \pi/3, \hspace{1cm}\:\:\: \gamma= \pi/9 \,\,\,.
\]
These processes give rise to the corresponding 
residue equations  
\be \label{doub}
\ba{c}
\bd
-i\lim_{\th\rightarrow i2\pi/3}(\th-i2\pi/3)F_{24}^{\Theta}(\th)=
\Gamma_{21}^1\Gamma_{41}^1\,F_{11}^{\Theta}(i\pi/3) \,\, ,
\ed
\\
\\
\bd
-i\lim_{\th\rightarrow i\pi/3}(\th-i\pi/3)F_{24}^{\Theta}(\th)=
\Gamma_{21}^3\Gamma_{41}^1\,F_{31}^{\Theta}(i\pi/9)\,\,\,.
\ed
\ea
\ee
which are indeed fulfilled. This example clearly shows the 
over--determined nature of the bootstrap equations and their 
internal consistency. 

The next FF in order of increasing value of the energy of the 
asymptotic state is given by the lightest $Z_2$ even three--particle state 
$\mid A_1 A_1 A_2\,\rangle$. The FF may be parametrized in the following way
\be
F_{112}^{\Theta}(\th_a,\th_b,\th_c)=
\frac{F_{11}^{min}(\th_{ab})\,F_{12}^{min}(\th_{ac})\,F_{12}^{min}(\th_{bc})}
{D_{11}(\th_{ab})\,D_{12}(\th_{ac})\,D_{12}(\th_{bc})}\,
\frac{Q_{112}^{\Theta}}{\cosh\th_{ac}+\cosh\th_{bc}} \,\, ,
\label{three}
\ee
where $F_{11}^{min}$ and $D_{11}^{min}$ are given by equations 
(\ref{f11min7}) and (\ref{D11min7}), while
\be
F_{12}^{min}(\th)= g_{13/18}(\th)\,\,g_{7/18}(\th) \,\,\, ,
\ee
and
\be
D_{12}(\th) = {\cal P}_{13/18}(\th) \,\,{\cal P}_{7/18}(\th)\,\,\,.
\ee
We have introduced into (\ref{three}) the term
\[
\frac{1}{\cosh\th_{ac}+\cosh\th_{bc}} \,\, ,
\]
to take into account the kinematical pole of this FF at $\th_a=\th_b+i\pi$.
The polynomial $Q_{112}$ in the numerator can be further decomposed as
\be
\label{q112pol}
Q_{112}^{\Theta}(\th_a,\th_b,\th_c)\,=\, P^2 \,P_{112}^{\Theta} \,\, ,
\ee
where $P^2$ is the kinematical polynomial expressed by 
\be
P^2 = 2m_1^2 + m_2^2 + 2m_1^2\cosh\th_{ab}
+2m_1 m_2(\cosh\th_{ac}+\cosh\th_{bc})\, .
\ee
The degree of $P_{112}^{\Theta}$ can be computed by means of the asymptotic 
behaviour in the three variables $\th_{a,b,c}$ separately. This gives
the following results for $Q\sim\exp\left[x_i\th_i\right]$:
\be
x_a=x_b=1\,\mbox{ and }\, x_c=2\, .
\ee
Hence, a useful parametrization of the polynomial $P_{112}$ is given by
\be
P_{112}^{\Theta}(\th_a,\th_b,\th_c)\,=\,p_0 + 
p_1\cosh\th_{ab}+p_2(\cosh\th_{ac}
+ \cosh\th_{bc}) + p_3\cosh\th_{ac}\cosh\th_{bc} \,\,\, ,
\ee
where four unknown constants have to be determined through the poles of 
$F_{112}^{\Theta}$. By using the kinematical pole at $\th_{ab}=i\pi$ and 
the bound state poles at $\th_{ab}= i\frac{5\pi}{9},\,\, i\frac{\pi}{9}$ and 
$\th_{ac}= i\frac{13\pi}{18},\,\, i\frac{7\pi}{18}$,
one obtains a redundant but nevertheless consistent system of five  equations 
in the four unknown $p_i$ whose solution 
is given by 
\be
\label{q112}
p_0 = -p_1 = \frac{p_3}{2} = -39.74991118... \,\,\,\,\, ,
\hspace{5mm}
p_2 = -198.2424080...
\ee
The other FFs which we have computed 
correspond to the states listed in Table 3. 
The values of the one--particle FFs
are collected in Table 4, while the results concerning the two--particle
computations are encoded in Table 5 via the coefficients $a_{ab}^{k}$ 
of the polynomials $P_{ab}(\th)$. 

\subsection{Recursive Equations of Form Factors in the TIM}
For sake of completeness, we now illustrate an efficient technique 
to compute multiparticle FFs. This is based on recursive 
identities which relate FFs of the type $F_{1,1,\ldots,1}$ with different 
(even) numbers of fundamental particles. Once these FFs are known, 
those relative to $Z_2$ even multi--particle state involving heavier particles 
may be obtained through bootstrap procedure. In general this way of proceeding
is the simplest one as far as FFs with three or more particles are 
concerned.
In order to write down these recursive equations, we can adopt the following 
pa\-ra\-meteri\-zation for the $2n$--particles FF $F_{1,1,\ldots,1}$:
\be
F_{1,1,\ldots,1}  (\th_1,\ldots,\th_{2n}) \equiv 
\F_{2n}  (\th_1,\ldots,\th_{2n})   
  =\,\frac{H_{2n} Q_{2n}(x_1,\ldots,x_{2n})}{\sigma_{2n}^{n-1}}
\prod_{i<k} \frac{F_{1\,1}^{min}(\th_{ik})}{D_{11}(\th_{ik})}
 \frac{1}{x_i + x_k} \,.
\ee
Here and in the following $\sigma_k (x_1,\ldots,x_{2n}) $ represents  
the symmetric polynomials of degree $k$ in the variables
$x_i= e^{\th_i}$ defined through their generating function
\be
\prod_{k=1}^{m} (x + x_k) = \sum_{j=0}^m  x^{m-j}  
\sigma_{j}(x_1,\ldots,x_m) \,\,\, .
\ee
$F_{11}$ and $D_{11}$ are defined by (\ref{f11min7}) and (\ref{D11min7})
while $H_n$ is an overall multiplicative constant and $Q_n$ is a symmetric 
polynomial  in its variables. The 
factors  $(x_i + x_k)^{-1}$ give a suitable parametrization of the 
kinematical poles, while the dynamical poles are taken into account by the
functions $D_{11}$'s.

The polynomial $Q_{2n}$ in the numerator can be factorized as  
\be
Q_{2n}(x_1,\ldots,x_{2n}) = \sigma_1 \sigma_{2n-1} P_{2n}(x_1,\ldots,x_{2n})
\,\,\, ,
\ee
since the FF will be proportional to the kinematical term $P^2$ relative 
to the total momentum which can be conveniently written as
\be
P^2 = m_1^2 \:\:\frac{\sigma_1\: \sigma_{2n-1}}{\sigma_{2n}} \,\,\,.
\ee
The Lorentz invariance of the FF requires $P_{2n}$ to be an 
homogeneous polynomial with respect to all the  $x_i$'s of total degree
\be 
\label{degE7}
\mbox{deg} \, P_{2n} = 4 n^2 - 5 n \,\, ,
\ee
while the condition (\ref{boundff}), knowing that $\Delta_{\epsilon} = 1/10$, 
imposes an upper bound to the degree in a single $x_i$, given by 
\be 
\label{degiE7}
\mbox{deg}_{x_i} \, P_{2n} < 4 n - 22/5  \,\,\,.
\ee
Writing down the most general expression of $P_{2n}$ 
as a symmetrical polynomial in the basis of the $\sigma_k$'s 
and taking into account the above conditions, one can determine 
the relative coefficients by means of the recursive equations. 
A first set of recursive relations is obtained by plugging 
the parametrization of $\F_{2n}$ into the equation of kinematical poles
(\ref{kinpole}); the polynomial $Q_n$ are then solution 
of the recursive equation
\be\label{recE7}
Q_{2n+2}(-x,x,x_1,\ldots,x_{2n}) = -i\: Q_{2n}(x_1,\ldots,x_{2n}) 
\:U_{2n}(x|x_i) ,
\ee
where the polynomial $U_{2n}$ is given by
\begin{eqnarray} 
U_{2n}(x|x_i) &=& \prod_{i=1}^n \prod_{\alpha \in {\cal A}_{1\,1}} 
(x + e^{-i \pi \alpha } x_i)
            (x - e^{i \pi\alpha } x_i) +\\
            & & -\prod_{i=1}^n \prod_{\alpha \in {\cal A}_{1\,1}} 
(x - e^{-i \pi\alpha } x_i)
            (x + e^{i \pi\alpha } x_i) \, .
            \nonumber 
\end{eqnarray} 
The overall constants $H_n$ have been fixed to be 
\be 
\label{HE7}
H_{2n} = 2 \,\pi\, m_1^2 \;\left( 16  
      \prod_{\alpha \in {\cal A}_{1\,1}} g_\alpha (0) 
      \frac{\cos^4(\pi\alpha/2)}{\sin(\pi\alpha)}
      \right)^{- n(n-1)} \,\, , 
\ee
with $H_2= 2 \pi m_1^2$. 
Given $Q_{2n}$, eq.\,(\ref{recE7}) restricts the form of the 
polynomial $Q_{2n+2}$, although these equations cannot determine 
uniquely all its coefficients. In fact, polynomials containing 
the kernel factor 
$ \prod_{i,j=1}^{2n} (x_i + x_j)
$ can be added to a given solution $Q_{2n+2}$ with an arbitrary multiplicative 
factor, without affecting the validity of eq.\,(\ref{recE7}). In order 
to have a more restrictive 
set of equations for the coefficients of the polynomials $Q_{2n}$, 
we employ the recursive equations (\ref{boundfpole}). 
To relate $\F_{2n+2}$ and $\F_{2n}$, we consider two successive fusions 
$A_1 A_1 \rightarrow A_2$ and $A_2 A_1 \rightarrow A_1$, obtaining the 
following equations 
\be
\label{recE7b}
Q_{2n+2}(-\varphi x,x,\varphi x,x_2,\ldots,x_{2n}) = 
\phi_n\: {\cal M} \: (\Gamma^2_{1\,1})^2\: x^5 \:
Q_{2n}(x,x_2,\ldots,x_{2n})\: P_{2n}(x|x_i)
\ee
where 
\[
{\cal M} = 4 \cos(5\pi /18) \cos(8\pi /18) , 
\] 
\[ 
\phi_n = (-1)^{n+1} \exp\Bigl(-i\pi(10n+1)/18\Bigr),
\]
\[
\varphi = \exp(-i 4 \pi/9) , 
\]
and
\be
P_{2n}(x|x_i) = \prod_{i=2}^{2n} (x - e^{i 8\pi/9 } x_i) (x - e^{i 5\pi/9 } x_i)(x + e^{i \pi/3 } x_i)
                 (x +  x_i) \,\,\,.
\ee
As an application of the above equations, let us consider the determination 
of the FF $\F_{4}$. Taking into account eqs. (\ref{degE7}) 
and (\ref{degiE7}), we can write the following general parametrization for 
$P_4$ as
\be
 P_4(x_1,\ldots,x_4) = c_1 \: \sigma_1^2 \sigma_4 +
                       c_2 \: \sigma_2 \sigma_4 +
                       c_3 \:\sigma_1\sigma_2\sigma_3 +
                       c_4 \:\sigma_3^2 +
                       c_5 \: \sigma_2^3 \,\,.
\ee
\mbox{}From (\ref{recE7}), knowing $Q_2 = \sigma_1$, one gets a first set of 
equations on the $c_i$'s
\begin{eqnarray}
c_2 &\!\!\! = &\!\!\! 4 \:\Bigl(\,2 \,\sin(\pi/9)\, + \,\sin(\pi/3)\,
              +\, 2 \,\sin(4 \pi/9)\, \Bigr),\nonumber  \\
c_5 &\!\!\! = &\!\!\! -4\: \Bigl(\, \sin(\pi /9)\, + \,
             \sin(4 \pi/9)\,\Bigr),\nonumber \\
c_4 &\!\!\! = &\!\!\! c_1,  \\
c_3 &\!\!\! = &\!\!\! c_5 - c_1 .\nonumber 
\end{eqnarray}
The residual freedom in the parameters reflects the presence 
of kernels of eq. (\ref{recE7}). Given any solution $Q_4^*$, 
the space of solutions is spanned by
\be
Q_4^\alpha = Q_4^* + 
\alpha\: \sigma_1 \:\sigma_3 \:\prod_{i,j=1}^{4} (x_i + x_j) , 
 \;\;\;\;\;\;\;\;\;\;\;\;\alpha\in \mbox{{\bf C}}
\ee
Eq. (\ref{recE7b}) solves this ambiguity giving the last needed equation
\be
c_1 = 2 \: \frac{4 \,\cos(\pi/18)\, -\, 11\, \cos(\pi/6)\,+\,12\,
      \cos(5 \pi/18) \,-\,       8\, \cos(7 \pi/18)}
   { 3 \,+ \,5 \,\cos(5 \pi/9)\, +\, \cos(\pi/3) \,- \,3 \,\cos(\pi/9) } \,\,.
\ee
Finally one directly computes $H_4$  from (\ref{HE7}).

The knowledge of $\F_4 = F_{1111}$ allows us to compute through 
successive applications of (\ref{boundfpole}) almost all the
FFs we needed in order to reach the required precision
of the FF expansion of the correlation function. We have used the obtained FFs
to compute the  two--point correlation function of $\Theta$ by means of the 
truncated spectral representation (\ref{formexp}). 
A plot of $\langle\,\Theta(x)\,\Theta(0)\,\rangle$
as a function of $|x|$ is drawn in Figure 5.
To control the accuracy of this result 
we have tested the fast convergence of the spectral series
on the checks relative to the first two moments of the correlation 
function eqs. (\ref{usum}) and (\ref{csum}); 
the single contributions of each multiparticle state
in the two series are listed in Table 3 and the partial sum 
is compared to the exact known values of the central charge $c$ and of 
the free energy amplitude $U$. 
A fast convergence behaviour of the spectral sum is indeed 
observed and therefore the leading dominant role of the first 
multiparticle states  in eq.(\ref{formexp}) is established.

\resection{Form Factors of the Energy Operator in the Thermal Deformed 
Tricritical Potts Model} \label{TPM}

In this section we will consider the FF computation 
for the Quantum Field Theory defined by the leading  thermal deformation of the
Tricritical 3--state Potts
Model (TPM).
Our strategy will resemble the one already applied to  the TIM, with 
suitable generalizations in order
to deal with this theory  of degenerate mass spectrum.

\subsection{Generalities of the TPM}
The 3--state Potts Model at its tricritical point  may be 
identified with the universality class of a subset of the minimal conformal 
model ${\cal M}_{6,7}$ \cite{BPZ}. Its central charge is $c=6/7$. The model 
is invariant under the permutation group $S_3$. The group $S_3$ is the 
semi--direct product of the two abelian groups $Z_2$ and $Z_3$, where the 
$Z_2$ group may be regarded as a charge conjugation symmetry implemented by 
the generator ${\cal C}$. For the generator $\Omega$ of the $Z_3$ symmetry, 
we have $\Omega^3=1$ and $\Omega \,{\cal C} = - {\cal C}\, \Omega$. The irreducible 
representations of $S_3$ could be either singlets, invariant with respect 
to $\Omega$ (${\cal C}$ even or ${\cal C}$ odd) or $Z_3$ charged 
doublets. 

The off--critical model we are interested in, is obtained by perturbing the 
fixed point action by means of the leading thermal operator $\epsilon(x)$ 
with conformal dimension $\Delta = 1/7$. This is a singlet field under both 
symmetries, ${\cal C}$ and $\Omega$. Hence, the discrete $S_3$ symmetry of 
the fixed point is still preserved away  from criticality and correspondingly 
the particle states organize into singlets or doublets. The scattering 
amplitudes of the massive excitations produced by the thermal deformation 
of the Tricritical Potts Model are nothing  but the minimal $S$--matrix 
elements of the Affine Toda Field Theory based on the root system of $E_6$ 
(they have been determined and discussed in references \cite{FZ,Sotkov} and 
can be found in Table 7). Poles occur at values $i\alpha
\pi$ with $\alpha$ a multiple of $1/12$, $12$ being the Coxeter number of the 
algebra $E_6$. The reason of the $E_6$ structure in the massive model is 
due both to the equivalent realization of the critical model in terms of the 
coset $(E_6)_1 \otimes (E_6)_1 / (E_6)_2$ and to the fact that the leading 
energy operator $\epsilon(x)$ is associated to the adjoint representation 
in the decompostion of the fields \cite{GKO}. Then, once again, one may 
apply the argument of references \cite{EY} to conclude that the massive 
theory inherits the $E_6$ symmetry of the fixed point. 

The exact mass spectrum consists in two doublets $(A_l,A_{\lb})$ 
and $(A_h,A_{\hb})$, together with two singlet particle states 
$A_L$ and $A_H$ \cite{FZ,Sotkov}. Their mass ratios are given by 
\begin{eqnarray}
m_l &=& m_{\overline l} = M(g)\,, \nonumber \\
m_L &=& 2 \,m_l \cos\frac{\pi}{4} = (1.41421..) \,m_l\,,\\
m_h &=& m_{\overline h} = 2\, m_l \cos\frac{\pi}{12} = (1.93185..) 
\,m_l\,,\nonumber\\
m_H &=& 2\, m_L\cos\frac{\pi}{12} = (2.73205..) \,m_l\,,\nonumber 
\end{eqnarray}
where the mass scale depends on $g$ as \cite{FateevTBA} 
\EQ
M(g) \,=\, {\cal{C}} \, g^{\frac{7}{12}} ,
\EN
and
\EQ
 {\cal{C}} \,=\, \left[ 
   4 \, \pi^2 \:  \gamma(\mbox{$\frac{4}{7}$}) \, \gamma(\mbox{$\frac{9}{14}$})\,\gamma(\mbox{$\frac{5}{7}$})
         \,\gamma(\mbox{$\frac{11}{14}$}) 
              \right]^{\frac{7}{24}}
                \frac{2 \,\Gamma(\frac{1}{4})\,\Gamma(\frac{13}{12}) }
{\Gamma(\frac{1}{2})\,
\Gamma(\frac{2}{3})\,\Gamma(\frac{7}{6})} = 3.746559718\ldots.
\EN
The above values of the masses are proportional to the components of the 
Perron--Frobenius eigenvector of the Cartan matrix of the exceptional algebra 
$E_6$ and therefore the particles may be associated to the dots of 
the Dynkin diagram (see Figure 6). Hence, they may be put in correspondence 
with the following representations of $E_6$ (identified by their dimensions) 
\EQ
\begin{array}{ccl}
A_l &\rightarrow & 27 \hs\hs,\\
A_{\overline l} &\rightarrow & \overline{27}\hs\hs,\\
A_L &\rightarrow & 78\hs\hs,\\
A_h &\rightarrow & 351\hs\hs,\\
A_{\overline h} &\rightarrow & \overline{351}\hs\hs,\\
A_H &\rightarrow & 2925\hs\hs.
\end{array}
\EN
By introducing the alternative 
notation 
\[
\begin{array}{ll}
A_l \,\rightarrow & A_1 \,\, ,\\
A_{\overline{l}} \,\rightarrow & \overline{A}_1 \,\, ,\\
A_h \, \rightarrow & A_2 \,\, ,\\
A_{\overline{h}} \, \rightarrow & \overline{A}_2 \,\, ,\\
A_L \,\rightarrow & B_1 \,\, ,\\
A_H \,\rightarrow & B_2 \,\,\, ,
\end{array}
\]
the bootstrap fusions of this model 
can be written in the following  compact way
 \cite{GM} 
\EQ
\begin{array}{lcl}
A_i\times A_i &=& \ov{A}_1+\ov{A}_2\\
A_i\times A_{i+1} &=& \ov{A}_1+\ov{A}_2\\
A_i\times \ov{A}_i &=& B_i\\
A_i\times \ov{A}_{i+1} &=& B_1+B_2\\
A_i\times B_i &=& A_1+A_2\\
\ov{A}_i\times B_i &=& A_1+A_2\\
A_i\times B_{i+1} &=& A_{i+1}\\
B_i\times B_i &=& B_1+B_2\\
B_i\times B_{i+1} &=& B_1+B_2 \hs\hs. 
\end{array}
\EN
It is easy to check that the above fusion rules are a subset of the tensor 
product decompositions of the above representations of $E_6$ \cite{Braden}.

\subsection{Form Factors of the TPM}
After a brief description of the model, let us turn our attention to the 
determination of the matrix elements of the leading energy operator 
$\epsilon(x)$. Our strategy will be similar to that employed in the case of the
TIM. For the TPM, however, we have a more stringent selection rule 
coming  from the $Z_3$ symmetry. Given the even parity of the operator 
$\epsilon(x)$ and its neutrality under the $Z_3$ symmetry, the only matrix 
elements which are different  from zero are those of singlet 
(multiparticle) states and they are the only contributions which enter 
the spectral representation series (\ref{formexp}). For convenience, 
the first such states ordered according to the
increasing value of the $s$--variable are listed in Table 8. Because 
of the selection rules, 
one very soon encounters three-- and four--particle 
states among the first contributions, and therefore, the 
computation of FFs becomes in general 
quite involved.

Let us briefly illustrate the most interesting FF computations of this 
model. As far as one-- and two--particles FFs are concerned, we just quote 
the result of the computations since they are quite straightforward and can 
be obtained by following the same strategy already adopted for the TIM; the 
one--particle FFs are given in Table 9, while the coefficients $a_{ab}^{k}$ 
of the polynomials $P_{ab}(\th)$ of eq.\,(\ref{ppol}) are listed in Table 10.
The need to compute several three--particle FFs suggests however to adopt
a more systematic technique based on the recursive structure of the FF.  
The lowest neutral mass state is given in this model by a doublet of 
conjugated particles $l$ and $\lb$. Hence, in order to build useful 
``fundamental" singlet multiparticle FFs we have to consider 
recursive equations relating FFs of the kind 
$\F_{n (l\,\lb)} \equiv F_{l\,\lb\,l\,\lb \ldots l\,\lb}$, with an arbitrary
number  of particle--antiparticle pairs. From the knowledge of 
$F_{l\,\lb\, l\,\lb}$ obtained as solutions of the recursive equations, 
we can next derive (by bootstrap fusion) all the three--particle
FFs we need in our determination of the correlation function. 
To write these recursive equations, let us parametrize the FFs as 
\begin{eqnarray} 
\label{fnllb}
\F_{n (l\,\lb)} (\beta_1,\bb_1,\ldots,\beta_n,\bb_n) &=&
 \frac{H_n \, Q_n(x_1,\xb_1,\ldots,x_n,\xb_n)}
{(\sigma_n \sigb_n)^{n-1}} \cdot \\
 & &  \!\!\!\!\!\!\!\!\!\!\!\!\!\!\!\!\!\!\!\!\!\!\!\!\!\!\!\!
 \!\!\!\!\!\!\!\!\!\!\!\!\!\!\!\!\left(
\prod_{1\leq i < k \leq n} 
\frac{F_{ll}^{min}(\beta_{ik}) \,\, F_{\lb\,\lb}^{min}(\bb_{ik})}
 {D_{ll}(\beta_{ik})\,\,\,\, D_{\lb\,\lb}(\bb_{ik})} 
\right)  \left(\prod_{r,s=1}^n 
\frac{\widehat{F_{l\lb}^{min}}(\beta_r - \bb_s)}{(x_r + \xb_s) \,\,D_{l\lb}(\beta_r - \bb_s)} \right),
\nonumber
\end{eqnarray}
where 
\be
\widehat{F_{l\lb}^{min}}(\beta_r - \bb_s) \equiv \left\{
\begin{array}{ll}
F_{l\lb}^{min}(\beta_r - \bb_s) & \;\;\mbox{if $r\leq s$} \,\, ,\\
&  \\
F_{\lb l}^{min}( \bb_s - \beta_r  ) & \;\;\mbox{otherwise} \,\,\,.
\end{array} \right.
\ee
In these expressions $x_i = e^{\beta_i}$  and
$\sigma_m$ is the symmetrical polynomial of degree $m$ in the $x_i$'s 
(the quantities $\xb_i$ and $\sigb_m$ are analogously defined in terms
of the $\bb_i$'s).
The two--particle minimal FFs are given by (see eqs. (\ref{fminE6}) and (\ref{dabE6}))
\begin{eqnarray}
& &  \!\!\!\!\!\!\!\!\!
\frac{F_{ll}^{min}(\beta)}{D_{ll}(\beta)} 
=\frac{ F_{\lb\,\lb}^{min}(\beta)}{D_{\lb\,\lb}(\beta)} = 
\frac{-i \sinh (\beta/2)\; h_{1/6}(\beta)\;h_{2/3}(\beta)\;h_{1/2}
(\beta)}{p_{1/6}(\beta)\;p_{2/3}(\beta)} \,\, , \\
 & & \nonumber \\
& & \!\!\!\!\!\!\!\!\!
\frac{F_{l \lb}^{min}(\beta)}{D_{l \lb}(\beta)} 
= \frac{F_{\lb l}^{min}(\beta)}{D_{\lb l}(\beta)} = 
\frac{ h_{5/6}(\beta)\;h_{1/3}(\beta)\;h_{1/2}(\beta)}{p_{1/2}(\beta)} \,\,\,.
\end{eqnarray}
In (\ref{fnllb}), $H_n$ is just a multiplicative overall factor and 
$Q_n$ is a polynomial in its arguments. The latter is the only unknown 
quantity which can be computed through the recursive equations. 
The function $Q_n$ must be a  
symmetrical polynomial both in the $x_i$'s and in the $\xb_i$'s separately. 
Furthermore, 
it must be symmetric under charge conjugation, i.e. under the simultaneous 
exchange $x_i \leftrightarrow \xb_i$ $(\forall i=1 \ldots n)$.
Hence, it can be parametrized in terms of products of $\sigma$'s
and $\sigb$'s with suitable coefficients in order to guarantee the 
self--conjugacy. The factor $P^2$ for this set of particles takes the form 
\be
P^2 = \frac{(\sigb_{n-1} \sigma_n + \sigma_{n-1} \sigb_n )
(\sigma_1 + \sigb_1)}
     {\sigma_n \sigb_n} \; m_l^2 \, ,
\ee 
and, correspondingly $Q_n$ will be factorized as
\be \label{qn}
Q_n (x_1,\xb_1,\ldots,x_n,\xb_n)=  
(\sigb_{n-1} \sigma_n + \sigma_{n-1} \sigb_n )
 (\sigma_1 + \sigb_1) P_n(x_1,\xb_1,\ldots,x_n,\xb_n) \, .
\ee 
The Lorentz invariance of the FF requires $P_n$ to be an homogeneous 
polynomial with respect to all the  $x$'s and $\xb$'s of total degree
\be 
\label{deg}
\mbox{deg} \, P_n = 3 n^2 - 4 n \,\, ,
\ee
while the condition (\ref{boundff}), knowing that $\Delta_\varphi = 1/7$, 
imposes the following upper bound for the degree in a single $x_i$ 
($\xb_i$) 
\be 
\label{degi}
\mbox{deg}_{x_i} \, P_n < 3 n - 74/21 \,\,\,. 
\ee
These conditions drastically restrict the possible form of the polynomials 
$Q_n$. 

Let us write down the form assumed by the kinematical recursive equations 
by using the parametrization (\ref{fnllb}) 
\be \label{recE6}
Q_{n+1} (-x,x,x_1,\xb_1,\ldots,x_n,\xb_n) = i\; x \;U_n(x|x_i,)\;
Q_{n} 
(x_1,\xb_1,\ldots,x_n,\xb_n) \,\, ,
\ee
where (here ${\cal A}_{l\,l}= \{ 1/6,2/3,1/2\}$)
\begin{eqnarray} 
U_n(x|x_i,\xb_i) &=& \prod_{i=1}^n \prod_{\alpha \in {\cal A}_{l\,l}} 
(x - e^{i \pi \alpha } \xb_i)
            (x - e^{i \pi(1-\alpha) } x_i) - \\
            & & \prod_{i=1}^n \prod_{\alpha \in {\cal A}_{l\,l}} 
(x - e^{-i \pi\alpha } \xb_i)
            (x - e^{-i \pi(1-\alpha) } x_i) \, .
            \nonumber 
\end{eqnarray}
The overall constant is explicitly given by:
\be
H_{n} = 2 \,\pi\, m_l^2 \;\left( 2 \tan^2(\pi/6) \tan^2(5 \pi/12) 
      \prod_{\alpha \in {\cal A}_{l\,l}} g_\alpha (0) 
\sin(\pi\alpha)\right)^{-\frac{n(n-1)}{2}} \,\,\, .
\ee
However, the equations (\ref{recE6}) are not in general sufficient to fix 
all the coefficients of $Q_{n+1}$. A more stringent constraint is 
obtained by using twice eq.\,(\ref{boundfpole}) in relation with 
the processes $l\,l \rightarrow \lb$ and  $\lb\,\lb \rightarrow \l$.
The final equations take a very simple form:
\begin{eqnarray} 
\label{recE6b}
Q_{n+1}(\eta\,\yb, \eta\, y,\etab\,\yb,\etab \,y,x_2,\xb_2,\ldots,x_n,\xb_n) &=& \\
& & \nonumber \\ 
  & &\!\!\!\!\!\!\!\!\!\!\!\!\!\!\!\!\!\!\!\!\!\!\!\!\!\!\!\!\!\!\!\!
\!\!\!\!\!\!\!
     \!\!\!\!\!\!\!\!\!\!\!\!\!\!\!\!\!\!\!\!\!\!\!\!\!\!\!\!\!\!\!\!
\!\!\!\!\!\!\!
     \!\!\!\!\!\!\!\!\!\!\!\!
 = - (\Gamma_{l\,l}^{\lb})^2 \; y\, \yb\; 
W_n(y,\yb,x_2,\xb_2,\ldots,x_n,\xb_n) \;
Q_{n}(y, \yb,x_2,\xb_2,\ldots,x_n,\xb_n) \,\,, \nonumber 
\end{eqnarray}
where $\eta= e^{i \pi/3}$ and 
\begin{eqnarray}
W_n(x_1,\xb_1,\ldots,x_n,\xb_n) &=& \\ 
& & \nonumber \\
& & \!\!\!\!\!\!\!\!\!\!\!\!\!\!\!\!\!\!\!\!\!\!\!\!\!\!\!
    \!\!\!\!\!\!\!\!\!\!\!\!\!\!\!\!\!\!\!\!\!\!\!\!\!\!\!\!
\!\!\!\!\!\!\!\!\!  = (x_1 + \xb_1) 
(\xb_1 - e^{\frac{7 \pi i}{6}} x_1)(\xb_1 - e^{\frac{-7 \pi i}{6}} x_1)
  (\xb_1 - e^{\frac{ \pi i}{2}} x_1)(\xb_1 - 
e^{\frac{- \pi i}{2}} x_1) \,\cdot \nonumber \\ 
& & \nonumber \\
& & \!\!\!\!\!\!\!\!\!\!\!\!\!\!\!\!\!\!\!\!\!\!\!\!\!\!\!
    \!\!\!\!\!\!\!\!\!\!\!\!\!\!\!\!\!\!\!\!\!\!\!\!\!\!\!\!
\!\!\!\!\!\!\!\!\!
   \cdot  \prod_{i=2}^n  (\xb_1 + x_i) (x_1 + \xb_i) 
         (\xb_1 - e^{\frac{5 \pi i}{6}} \xb_i)(\xb_1 - 
e^{\frac{-5 \pi i}{6}} \xb_i)
         (x_1 - e^{\frac{5 \pi i}{6}} x_i)(x_1 - 
e^{\frac{-5 \pi i}{6}} x_i)  \nonumber \,\,.
\end{eqnarray}
Let us now illustrate how this procedure works in the case of $\F_{2 (l\,\lb)}$.
Let us start from $F_{l\,\lb}$; using eq.\,(\ref{norm}) we easily 
obtain $Q_1=1$ and $H_1= 2\pi \,m_l^2$. From eqs. (\ref{deg})
 and (\ref{degi}), the  
general parametrization  for $P_2$ is given by 
\begin{eqnarray}
P_2(x_1,\xb_1,x_2,\xb_2) &=& c_1 \, (\sigma_2^2 + \sigb_2^2) + c_2 \,
(\sigma_1\sigma_2 \sigb_1 +\sigb_1\sigb_2 \sigma_1)+ \\
& & \nonumber \\
& & + c_3 \,(\sigma_1^2\sigb_2  +\sigb_1^2\sigma_2 ) + 
c_4 \,\sigma_1^2\sigb_1^2 + c_5 \sigma_2\sigb_2 \nonumber \,\,.
\end{eqnarray}
Equation (\ref{recE6}) gives four equations for the five parameters
\begin{eqnarray}
c_4\!\!\! &=&\!\!\! - (3 + \sqrt{3}) \nonumber,\\
c_2 - c_3 \!\!\!&=&\!\!\! -3\,(2 + \sqrt{3})\nonumber,\\
c_1 - c_2 \!\!\!&=&\!\!\! 3 + 2\sqrt{3},\\
2 \,c_2 + c_5\!\!\! &=&\!\!\! -18 - 10\sqrt{3} \nonumber,
\end{eqnarray}
while eq. (\ref{recE6b}) solve the residual freedom yielding
\begin{eqnarray}
c_1 \!\!\!&=&\!\!\! - \frac{9 + 5 \sqrt{3}}{2} \nonumber,\\
c_2 \!\!\!&=&\!\!\! - \frac{3(5 + 3 \sqrt{3})}{2}\nonumber,\\
c_3 \!\!\!&=&\!\!\! - \frac{3(1 +  \sqrt{3})}{2},\\
c_4 \!\!\!&=&\!\!\! c_5 = - (3 + \sqrt{3}) \nonumber \,\,.
\end{eqnarray}
Once we have determined $H_1$ and $P_2$, we can obtain
$\F_{2 (l\,\lb)}$   from eqs. (\ref{fnllb}) and (\ref{qn}). From 
this four--particles FF it is also easy  to obtain
the three--particles FFs $F_{l\, l\, l}$,
$F_{l\, \lb\, L}$, $F_{l\, l\, h}$
applying the residue equation (\ref{boundfpolebis})  at 
the fusion angles $u_{\lb\,\lb}^{l}$, $u_{l\,\lb}^{L}$  and $u_{\lb\,\lb}^{h}$  
respectively. 
The explicit expressions of these three--particle FFs 
are given in Appendix B.

The FFs calculated for the TPM can be used to estimate the two--point function 
of the stress--energy tensor whose plot  is shown in Figure 7.  
The convergence of the series may be checked through the sum--rule tests: 
the contributions of each multiparticle state are listed in Table 8 
where the exact and computed values of $c$ and $U$ are 
compared. A very fast convergence behaviour is indeed observed which 
supports the validity of the spectral approach to  correlations functions
in integrable massive models.

\resection{Conclusions}

In this paper we have applied the Form Factor approach to estimate 
the correlation functions of the stress--energy tensor in the 
Tricritical Ising and 3--state Potts models. Both models have been 
perturbed away from the critical point by means of the leading 
thermal operator. In our computation, an important role has been 
played by the discrete symmetries of the two models, a $Z_2$ symmetry 
for the TIM and a $S_3$ symmetry for the TPM. These symmetries have 
in fact selected the appropriate particle states entering the spectral 
representation of the two--point correlator of the stress--energy 
tensor. This correlator has been plotted in Figure 5 for the TIM and 
in Figure 7 for the TPM: these plots are expected to be extremely 
precise in the large distance region ($m R \geq 1$) and sufficiently 
accurate in the crossover and ultraviolet regions ($m R \leq 1$). Obviously, 
a definite confirmation of their validity can only be obtained by 
a comparison with some experimental data or numerical simulations. 

\vspace{3mm}
{\em Acknowledgments}. It is a great pleasure to thank G. Delfino for 
useful discussions and helpful comments.

\newpage

\appendix
\appsection
In this appendix we collect some different explicit representations 
of the functions $g_\alpha(\th)$  and $h_\alpha(\th)$ together with 
some useful functional relations. 

Let us start by considering the non--degenerate field theories. In this 
case, the basic functions $g_{\alpha}$ needed to build the minimal form 
factors are obtained as solution of the equations  
\be
\label{ming}
\ba{l}
\bd
g_{\alpha}(\th)=-f_{\alpha}(\th)\,g_{\alpha}(-\th)\,\, ,
\ed
\\
\\
\bd
g_{\alpha}(i\pi+\th)= g_{\alpha}(i\pi-\th) \,\, ,
\ed
\ea                                                                           
\ee
where
\be \label{EFFEALFA}
f_{\alpha}(\th) = \frac{\tanh\frac{1}{2}\left(\th + i \pi \alpha\right)}
{\tanh\frac{1}{2}\left(\th - i \pi \alpha\right)}\,\,\,.
\ee
They are called minimal solutions because they do not present neither poles nor zeros
in the strip $Im \th \in (0,2\pi)$. They admit several equivalent 
representations. The first is the integral representation given by 
\be
\label{gmin}
g_{\alpha}=
\exp\left[2\int_0^{\infty}\frac{\de t}{t}
\frac{\cosh\left[(\alpha-1/2)t\right]}{\cosh t/2 \sinh t}
\sin^2(\hat{\th}t/2\pi)\right]\, ,
\ee
where $\hat{\th}=i\pi-\th$. The analytic continuation of the above expression 
is provided by the infinite product representation
\be
g_{\alpha}(\theta)= \prod_{k=0}^{\infty}
\left[ \frac{
\left[1 + \left(\frac{\hat\th/2\pi}{k+1 -\frac{\alpha}{2}}\right)^2
\right] \left[1 + \left(\frac{\hat\th/2\pi}{k+\frac{1}{2} + \frac{\alpha}{2}}
\right)^2\right] }
{\left[1 + \left(\frac{\hat\th/2\pi}{k +1 + \frac{\alpha}{2}}
\right)^2\right] \left[1 + \left(\frac{\hat\th/2\pi}{k+
\frac{3}{2} - \frac{\alpha}{2}}\right)^2
\right]}\right]^{k+1}
\, ,
\ee
which  explicitly  shows the position of the infinite number of poles outside 
the strip $Im \th \in (0,2\pi)$. Another useful representation particularly 
suitable for deriving functional equations is the following:
\be
g_\alpha(\th) = \prod_{k=0}^{\infty} 
\frac{\Gamma^2\left(\frac{1}{2} + k + \frac{\alpha}{2}\right)
\Gamma^2\left(1+k-\frac{\alpha}{2}\right)} 
{\Gamma^2\left(\frac{3}{2} + k - \frac{\alpha}{2}\right)
\Gamma^2\left(1+k+\frac{\alpha}{2}\right) }
\left|\frac{\Gamma\left(1+k+\frac{\alpha}{2} + i \frac{\hat\th}{2\pi}\right) 
\Gamma\left(\frac{3}{2} + k -\frac{\alpha}{2} +i \frac{\hat\th}{2\pi}\right)}
{\Gamma\left(1+k-\frac{\alpha}{2} + i \frac{\hat\th}{2\pi}\right) 
\Gamma\left(\frac{1}{2} + k +\frac{\alpha}{2} + i \frac{\hat\th}{2\pi}\right)}
\right|^2 \,\, ,
\label{Gamma}
\ee
where we have used the notation 
\be
\left| \Gamma (a + i  \hat{\th}/2\pi)\right|^2 \equiv
\Gamma (a + i  \hat{\th}/2\pi)\,\,\Gamma (a -i  \hat{\th}/2\pi) \: .
\nonumber
\ee
A representation that is particularly suitable for 
numerical evaluations is the mixed one 
\be \label{mixed}
\begin{array}{rcl} 
   g_{\alpha}(\th) &=&\bd
\prod_{k=0}^{N-1}
\left[ \frac{
\left[1 + \left(\frac{\hat\th/2\pi}{k+1 -\frac{\alpha}{2}}\right)^2
\right] \left[1 + \left(\frac{\hat\th/2\pi}{k+\frac{1}{2} + \frac{\alpha}{2}}
\right)^2\right] }
{\left[1 + \left(\frac{\hat\th/2\pi}{k +1 + \frac{\alpha}{2}}
\right)^2\right] \left[1 + \left(\frac{\hat\th/2\pi}{k+
\frac{3}{2} - \frac{\alpha}{2}}\right)^2
\right]}\right]^{k+1} \ed \times \\
& & \\
& \times& \bd   \exp\left[ 2 \int_0^{\infty} \frac{dt}{t}
\frac{\cosh\left[\frac{t}{2} (1 - 2\alpha)\right]}{\cosh\frac{t}{2} \sinh t}
(N + 1 - N e^{-2t}) e^{-2 N t} \sin^2\frac{\hat\th t}{2 \pi}\right] \ed\,\,\,.
\end{array}
\ee
In this formula $N$ is an arbitrary integer number which may be adopted to 
obtain a fast convergence of the integral. 

Using the integral representation (\ref{gmin}),  it is easy  to establish 
the asymptotic behaviour of $g_\alpha$ 
\be
\label{gas}
g_{\alpha}(\th)\sim e^{|\th|/2}\, \hspace{2cm}\mbox{for}\hspace{2cm}\,\th\rightarrow\infty\, .
\ee
The function $g_\alpha$ is normalized according to
\be
g_{\alpha}(i\pi) = 1 \,\,\, ,
\ee
and satisfies 
\be
g_{\alpha}(\th) = g_{1-\alpha}(\th) \,\,\, ,
\ee
with 
\be
g_0(\th) = g_1(\th) = - i \sinh\frac{\th}{2} \,\,\, .
\ee
The above functions satisfy the following set of functional equations  
\be
g_{\alpha}(\th + i \pi) g_{\alpha}(\th) =
-i \frac{g_{\alpha}(0)}{\sin\pi \alpha} (\sinh\th + i \sin\pi \alpha)
\,\,\, ,
\label{pinchkin}
\ee
\be
g_{\alpha}(\th + i\pi \gamma) g_{\alpha}(\th - i\pi \gamma) =
\left(\frac{g_{\alpha}(i \pi \gamma) g_{\alpha}(-i\pi \gamma)}
{g_{\alpha+\gamma}(0) g_{\alpha - \gamma}(0)}\right)
g_{\alpha + \gamma}(\th) g_{\alpha - \gamma}(\th) \,\,\, ,
\label{pinchbound}
\ee
\be
g_{1-\alpha}(\th) g_{\alpha-1}(\th) =
\frac{\sinh\frac{1}{2}[\th - i (\alpha-1) \pi] \sinh\frac{1}{2}[\th + i
(\alpha+1) \pi]}
{\sin^2\frac{\pi \alpha}{2}} \,\,\, .
\label{cancellation}
\ee
Let us turn our attention to the field theories with a degenerate mass 
spectrum. In complete analogy with the previous 
case, we start our analysis  from the minimal solutions of the equations
\be
\label{minh}
\ba{l}
\bd
h_{\alpha}(\th)=-s_{\alpha}(\th)\,h_{\alpha}(-\th)
\ed
\\
\\
\bd
h_{\alpha}(i\pi+\th)= h_{\alpha}(i\pi-\th) ,
\ed
\ea                                                                           
\ee
where 
\be \label{ESSEALFA}
s_{\alpha}(\th) = \frac{\sinh\frac{1}{2}\left(\th + i \pi \alpha\right)}
{\sinh\frac{1}{2}\left(\th - i \pi \alpha\right)}\,\,\,.
\ee
The function $h_\alpha (\th)$ is explicitly given in terms of 
the following equivalent representations
\be
\label{hmin}
h_{\alpha}(\theta)=
\exp\left[2\int_0^{\infty}\frac{\de t}{t} \,\,
\frac{\sinh\left[(1-\alpha)t\right]}{ \sinh^2 t} \,
\sin^2(\hat{\th}t/2\pi)\right]\, ,
\ee
\be
h_{\alpha}(\theta)= \prod_{k=0}^{\infty} 
\left(  \frac{1 + \left(\frac{\frac{\hat{\th}}{2 \pi}}
             {n + \frac{1}{2}+ \frac{\alpha}{2}}  \right)^2}
             {1 + \left(\frac{\frac{\hat{\th}}{2 \pi}}
             {n + \frac{3}{2}- \frac{\alpha}{2}}  \right)^2}
\right)^{k+1}\, ,
\ee
\be \bd
h_{\alpha}(\theta)= \prod_{k=0}^{\infty} 
\frac{\Gamma^2 (k +\frac{1}{2} +\frac{\alpha}{2}) \Gamma (k +1  - \frac{\alpha}{2}- \frac{i\th}{2\pi})
        \Gamma (k +2  - \frac{\alpha}{2}+ \frac{i\th}{2\pi})}
{\Gamma^2 (k +\frac{3}{2} -\frac{\alpha}{2}) \Gamma (k + \frac{\alpha}{2}- \frac{i\th}{2\pi})
        \Gamma (k + 1  + \frac{\alpha}{2}+ \frac{i\th}{2\pi})} \,\,.
\ed\ee
The mixed representation is in this case
\begin{eqnarray}
h_{\alpha}(\theta) \!\!\!&=\!\!\!&  \prod_{k=0}^{N+1} 
\left(  \frac{1 + \left(\frac{\frac{\hat{\th}}{2 \pi}}
             {n + \frac{1}{2}+ \frac{\alpha}{2}}  \right)^2}
             {1 + \left(\frac{\frac{\hat{\th}}{2 \pi}}
             {n + \frac{3}{2}- \frac{\alpha}{2}}  \right)^2}
\right)^{k+1} \times \\ 
\!\!\!& \!\!\!& \times \,\,\exp\left[2\int_0^{\infty}\frac{\de t}{t}\,
(N+1-N \, e^{-2 t})\, e^{-2Nt} \,\,
\frac{\sinh\left[(1-\alpha)t\right]}{ \sinh^2 t}\,
\sin^2(\hat{\th}t/2\pi)\right]\,\,\, , \nonumber
\end{eqnarray}
and the asymptotic behaviour  depends on the value of $\alpha$
\be
\label{has}
h_{\alpha}(\th)\sim e^{\frac{(1-\alpha)|\th|}{2}}\,
\mbox{ for }\,\th\rightarrow\infty\, .
\ee
The function $h_\alpha$ is normalized according to 
\be
h_{\alpha}(i\pi) =1
\ee
and satisfies the following functional equations: 
\be
\bd
\begin{array}{l}
h_{\alpha}(2 \pi i- \theta) = h_{\alpha}(\theta) , \\
\\
h_0(\th) = - i \sinh(\th/2), \\
\\
h_1(\th) = 1, \\ 
\\
h_{1+\alpha}(\theta) = h^{-1}_{1-\alpha}(\theta)  , 
\end{array}
\ed
\ee
The basic  ``composition rules'' for products of $h_\alpha$'s are:
\be
\bd
\begin{array}{c}
h_{\alpha}(\theta) \: h_{-\alpha}(\theta) = {\cal P}_\alpha(\theta) ,\\
\\
\bd h_{\alpha}(\theta + i \pi\gamma)\: h_{\alpha}(\theta - i \pi\gamma) =
\frac{h_{\alpha}( i \pi\gamma)\: h_{\alpha}( - i \pi\gamma)}
 {h_{\alpha +\gamma}(0) \: h_{\alpha - \gamma}(0)}\:
 h_{\alpha +\gamma}(\theta) \: h_{\alpha - \gamma}(\theta) \ed \\
 \\ 
 \bd
h_{\alpha}(\theta + i\pi) \: h_{1-\alpha}(\theta ) = 
\frac{h_{1-\alpha}(0)}{\cosh(\frac{i\pi\alpha}{2})} \:
\cosh\frac{\theta - i\pi\alpha}{2} \ed
\end{array}
\ed
\ee
where the polynomial ${\cal P}$ is defined in (\ref{pmin}) 
of Section 2. 

Finally, since $f_\alpha (\th) = 
s_\alpha (\th)s_{1-\alpha} (\th)$, the function $g_\alpha$ 
can be obtained  from the $h_\alpha$'s simply through:  
\be 
g_{\alpha}(\theta)=
h_{\alpha}(\theta) \: h_{1-\alpha}(\theta)  .\\
\\
\ee

\appsection
In this appendix we briefly report the results of the 
three--particle FFs relevant for our computation in the TPM. 
These FFs have been derived by 
applying the residue equations (\ref{boundfpolebis})
to the four--particles FF $F_{l\,\lb\,l\,\lb}^\Theta$, 
as explained in section \ref{TPM}.
In writing their final form, we have extensively used the formulas reported 
in Appendix A. 
The two--particle minimal 
FFs $F^{min}_{ab}$ appearing in the 
expressions which follow are defined by eq. (\ref{fminE6}) while 
the $D_{ab}$ factors parametrizing the dynamical poles are 
defined by eq. (\ref{dabE6}).

The FF $F^\Theta_{l\,l\,l}$ is obtained  from $F_{l\,\lb\,l\,\lb}^\Theta$
through the residue equation at $u_{\lb\,\lb}^{l}= 2 i\pi /3$
\EQ
F^\Theta_{l\,l\,l}(\th_1,\th_2,\th_3) = 
\left(
\prod_{i<j} \frac{F_{l\,l}^{min}(\th_{ij})}{D_{l\,l}(\th_{ij})} 
\right) \;
\left( 3 \, m_l^2 + 2\, m_l^2 \sum_{i<j} \cosh (\th_{ij}) \right)\; a_{l\,l\,l}^0 \; .
\EN
In this expression one immediately recognizes the ``minimal'' part, 
the dynamical poles and the $P^2$ polynomial, while the only remaining 
polynomial in the $\cosh(\th_{ij})$'s allowed by eq.\,(\ref{boundff}) 
is simply a constant given by
\[
a_{l\,l\,l}^0 = -102.3375342 \ldots\, .
\]
The FF $F^\Theta_{l\,\lb\,L}$, is  obtained   
from $F_{l\,\lb\,l\,\lb}^\Theta$ by using 
eq.(\ref{boundfpolebis}), with $u_{\lb\,l}^{L}= i \pi /2$. Its final 
expression is given by 
\begin{eqnarray}
F^\Theta_{l\,\lb\,L}(\th_1,\th_2,\th_3) \!\!\!&=\!\!\!& 
   \frac{F^{min}_{l\,\lb}(\th_{12})\, F^{min}_{l\,L}(\th_{13})\,F^{min}_{\lb\,L}(\th_{23}) }
        {D_{l\,\lb}(\th_{12})\, D_{l\,L}(\th_{13})\, D_{\lb\,L}(\th_{23})} \cdot \nonumber\\ & & \nonumber\\
\!\!\!& \!\!\!&   \!\!\!\!\!\!\!\!\!\!\!\!\!\!\!\!\!\!\!\!\!      
 \cdot \frac{2\,m_l^2 + m_L^2 + 2 m_l^2 \cosh(\th_{12}) + 2 m_l \,m_L \Bigl(\cosh(\th_{13}) + \cosh(\th_{23}) \Bigr) }
        {\cosh(\th_{13}) + \cosh(\th_{23})} \cdot \\ & & \nonumber\\
       \!\!\!& \!\!\!& \!\!\!\!\!\!\!\!\!\!\!\!\!\!\!\!\!\!\!\!\!
     \cdot  \biggl(
        a_{l\,\lb\,L}^0 \,\Bigl(1-\cosh(\th_{12}) + 2\,\cosh(\th_{13})\,\cosh(\th_{23})\Bigr) +
        a_{l\,\lb\,L}^1 \Bigl(\cosh(\th_{13}) + \cosh(\th_{23})\Bigr)
       \biggr) \nonumber \;.
\end{eqnarray}
This expression also exhibits a kinematical pole due to the presence
of a particle--anti\-par\-ticle pair $l\,\lb$. Moreover there is a nontrivial 
polynomial in the $\cosh(\th_{ij})$'s with coefficients given by
\[
a_{l\,\lb\,L}^0 = -70.50661963 \ldots\, ,
\]
\[
a_{l\,\lb\,L}^1 = -235.9197474\ldots \, .
\]
Finally, applying eq.(\ref{boundfpolebis}) to $F_{l\,\lb\,l\,\lb}^\Theta$
at $u_{\lb\,\lb}^{h}= i \pi /6$ one obtains 
\begin{eqnarray}
F^\Theta_{l\,l\,h}(\th_1,\th_2,\th_3) \!\!\!&=\!\!\!&
 \frac{F^{min}_{l\,l}(\th_{12})\, F^{min}_{l\,h}(\th_{13})\,F^{min}_{l\,h}(\th_{23}) }
        {D_{l\,l}(\th_{12})\, D_{l\,h}(\th_{13})\, D_{l\,h}(\th_{23})} \cdot  \\ & & \nonumber\\
        \!\!\!& \!\!\!&   \!\!\!\!\!\!\!\!\!\!\!\!\!\!\!\!\!\!\!\!\!\!\!\!\!\cdot
\biggl(2\,m_l^2 + m_h^2 + 2 m_l^2 \cosh(\th_{12}) + 2 m_l \,m_h \Bigl(\cosh(\th_{13}) + \cosh(\th_{23}) \Bigr)
\biggr)\cdot\nonumber\\ & & \nonumber\\
       \!\!\!& \!\!\!& \!\!\!\!\!\!\!\!\!\!\!\!\!\!\!\!\!\!\!\!\!\!\!\!\!\cdot
\biggl(
a_{l\,l\,h}^0 + a_{l\,l\,h}^1\, \Bigl(\cosh(\th_{13}) + \cosh(\th_{23})\Bigr)
 + a_{l\,l\,h}^2\, \cosh(\th_{12}) + a_{l\,l\,h}^3\, \cosh(\th_{13})\,\cosh(\th_{23})
\biggr) \nonumber
\end{eqnarray}
where the coefficients $a_{l\,l\,h}^k$ are given by 
\[
a_{l\,l\,h}^0 =   78134.00044 \ldots\, ,
\]
\[
a_{l\,l\,h}^1 = 72661.45729 \ldots\, ,
\]
\[
a_{l\,l\,h}^2 =  31793.68905 \ldots\, ,
\]
\[
a_{l\,l\,h}^3 = 43430.98692 \ldots\, .
\]


\newpage

\vspace{25mm}

{\bf Table Captions}

\vspace{1cm}

\begin{description}
\item [Table 1] Particle spectrum, mass ratios and $Z_2$--charges in the TIM. 
\item [Table 2] Two--particle $S$--matrix elements of the TIM; the notation
$(x)\equiv f_{x/h}(\th)$ has been followed, where $h=18$ is the Coxeter number of  $E_7$ and
the function $f_\alpha$ is defined in eq. (\ref{EFFEALFA}). Superscripts label the particles occurring as
bound states at the fusion angles $u_{ab}^c= x\pi/h$.
\item [Table 3]  The first $Z_2$--even multiparticle states 
of the TIM ordered according to the increasing value
 of the center--of--mass energy and their relative contributions to the spectral sum 
 rules of the central charge $c$ and  the free--energy amplitude $U$ . 
\item [Table 4] One--particle FFs of  the $Z_2$--even particles of the TIM.
\item [Table 5] Coefficients which enter  eq. (\ref{ppol}) for the lightest two--particle
FFs of the TIM. 
\item [Table 6] Particle spectrum, mass ratios and $Z_3$--charges in the TPM. 
\item [Table 7] Two--particle $S$--matrix elements of the TPM. 
In this case $[x]\equiv s_{x/h}(\th)$,
 where $s_\alpha$ is defined in eq. (\ref{ESSEALFA}) 
 and $h=12$ is the Coxeter number of  $E_6$.
\item [Table 8]The first $Z_3$--neutral multiparticle states of the TIM ordered 
 according to the increasing value
 of the center--of--mass energy and their relative contributions to the spectral sum 
 rules of the central charge $c$ and  the free--energy amplitude $U$ .  
\item [Table 9] One--particle FFs of  the $Z_2$--neutral particles of the TPM. 
\item [Table 10] Coefficients  which enter eq. (\ref{ppol}) for the lightest two--particle
FFs of the TPM.
\end{description}

\newpage


\begin{center}
\begin{tabular}{||c|c|c||} \hline
{\it particle}\rule[-2mm]{0mm}{7mm} & \multicolumn{1}{c|}{ $mass/m_1$}  &
\multicolumn{1}{c||}{$Z_2$ {\it charge}}   \\ \hline\hline
\rule[-2mm]{0mm}{7mm}$A_{1}$ &1.00000 & $-1$  \\ \hline
\rule[-2mm]{0mm}{7mm}$A_{2}$ &1.28558 & $1$  \\ \hline
\rule[-2mm]{0mm}{7mm}$A_{3}$ &1.87939 & $-1$  \\ \hline
\rule[-2mm]{0mm}{7mm}$A_{4}$ &1.96962 & $1$  \\ \hline
\rule[-2mm]{0mm}{7mm}$A_{5}$ &2.53209 & $1$  \\ \hline
\rule[-2mm]{0mm}{7mm}$A_{6}$ &2.87939 & $-1$  \\ \hline
\rule[-2mm]{0mm}{7mm}$A_{7}$ &3.70167 & $1$  \\ \hline
\end{tabular}
\end{center}
\begin{center}
{\bf Table 1}
\end{center}

\newpage

\begin{center}
\begin{tabular}{|c|c|}\hline
$a$ \,\, $b$ &
$S_{ab}$ \\ \hline \hline
\rule[-2mm]{0mm}{10mm}1 \,\, 1 &
$ -  \st{\bf 2}{(10)} \, \st{\bf 4}{(2)}  $\\ \hline
\rule[-2mm]{0mm}{10mm}1 \,\, 2 &
$ \st{\bf 1}{(13)} \, \st{\bf 3}{(7)} $\\ \hline
\rule[-2mm]{0mm}{10mm}1 \,\, 3 &
$ - \st{\bf 2}{(14)} \, \st{\bf 4}{(10)} \, \st{\bf 5}{(6)} $ \\ \hline
\rule[-2mm]{0mm}{10mm}1 \,\, 4 &
$  \st{\bf 1}{(17)} \, \st{\bf 3}{(11)} \, \st{\bf 6}{(3)} \, (9) $ \\ \hline
\rule[-2mm]{0mm}{10mm}1 \,\, 5 &
$  \st{\bf 3}{(14)} \, \st{\bf 6}{(8)} \,  (6)^2  $ \\ \hline
\rule[-2mm]{0mm}{10mm}1 \,\, 6 &
$ - \st{\bf 4}{(16)} \, \st{\bf 5}{(12)} \, \st{\bf 7}{(4)} \,  (10)^2  $ \\ \hline
\rule[-2mm]{0mm}{10mm}1 \,\, 7 &
$  \st{\bf 6}{(15)}  (9) \, (5)^2 \,  (7)^2  $ \\ \hline
\rule[-2mm]{0mm}{10mm}2 \,\, 2 &
$  \st{\bf 2}{(12)} \, \st{\bf 4}{(8)} \, \st{\bf 5}{(2)} $ \\ \hline
\rule[-2mm]{0mm}{10mm}2 \,\, 3 &
$  \st{\bf 1}{(15)} \, \st{\bf 3}{(11)} \, \st{\bf 6}{(5)} \, (9) $ \\ \hline
\rule[-2mm]{0mm}{10mm}2 \,\, 4 &
$  \st{\bf 2}{(14)} \, \st{\bf 5}{(8)} \,  (6)^2 $ \\ \hline
\rule[-2mm]{0mm}{10mm}2 \,\, 5 &
$  \st{\bf 2}{(17)} \, \st{\bf 4}{(13)} \, \st{\bf 7}{(3)}\,  (7)^2  \, (9) $ \\ \hline
\rule[-2mm]{0mm}{10mm}2 \,\, 6 &
$  \st{\bf 3}{(15)} \,  (7)^2  \,(5)^2  \, (9) $ \\ \hline
\rule[-2mm]{0mm}{10mm}2 \,\, 7 &
$  \st{\bf 5}{(16)} \, \st{\bf 7}{(10)^3} \, (4)^2  \,(6)^2   $ \\ \hline
\rule[-2mm]{0mm}{10mm}3 \,\, 3 &
$  - \st{\bf 2}{(14)} \, \st{\bf 7}{(2)} \, (8)^2  \,(12)^2   $ \\ \hline
\rule[-2mm]{0mm}{10mm}3 \,\, 4 &
$  \st{\bf 1}{(15)}  \, (5)^2  \,(7)^2 \,(9)  $ \\ \hline
\rule[-2mm]{0mm}{10mm}3 \,\, 5 &
$  \st{\bf 1}{(16)} \, \st{\bf 6}{(10)^3} \, (4)^2  \,(6)^2   $ \\ \hline
\rule[-2mm]{0mm}{10mm}3 \,\, 6 &
$  - \st{\bf 2}{(16)} \, \st{\bf 5}{(12)^3}\, \st{\bf 7}{(8)^3} \, (4)^2     $ \\ \hline
\rule[-2mm]{0mm}{10mm}3 \,\, 7 &
$   \st{\bf 3}{(17)} \, \st{\bf 6}{(13)^3}  \, (3)^2 \, (7)^4 \, (9)^2    $ \\ \hline
\end{tabular}
\end{center}
\begin{center}
{\bf Table 2} (Continued)
\end{center}
\newpage

\begin{center}
\begin{tabular}{|c|c|}\hline
$a$ \,\, $b$ &
$S_{ab}$ \\ \hline \hline
\rule[-2mm]{0mm}{10mm}4 \,\, 4 &
$   \st{\bf 4}{(12)} \, \st{\bf 5}{(10)^3}  \,\st{\bf 4}{(7)}  \, (2)^2     $ \\ \hline
\rule[-2mm]{0mm}{10mm}4 \,\, 5 &
$   \st{\bf 2}{(15)} \, \st{\bf 4}{(13)^3}  \,\st{\bf 7}{(7)^3}  \, (9)     $ \\ \hline
\rule[-2mm]{0mm}{10mm}4 \,\, 6 &
$   \st{\bf 1}{(17)} \, \st{\bf 6}{(11)^3}  \, (3)^2 \, (5)^2  \, (9)^2     $ \\ \hline
\rule[-2mm]{0mm}{10mm}4 \,\, 7 &
$   \st{\bf 4}{(16)} \, \st{\bf 5}{(14)^3}  \, (6)^4 \, (8)^4       $ \\ \hline
\rule[-2mm]{0mm}{10mm}5 \,\, 5 &
$   \st{\bf 5}{(12)^3} \, (2)^2  \, (4)^2 \, (8)^4       $ \\ \hline
\rule[-2mm]{0mm}{10mm}5 \,\, 6 &
$   \st{\bf 1}{(16)}\, \st{\bf 3}{(14)^3} \, (6)^4 \, (8)^4      $ \\ \hline
\rule[-2mm]{0mm}{10mm}5 \,\, 7 &
$   \st{\bf 2}{(17)}\, \st{\bf 4}{(15)^3} \,\st{\bf 7}{(11)^5} \, (5)^4 \, (9)^3      $ \\ \hline
\rule[-2mm]{0mm}{10mm}6 \,\, 6 &
$  - \st{\bf 4}{(14)^3} \,\st{\bf 7}{(10)^5} \, (12)^4 \, (16)^2      $ \\ \hline
\rule[-2mm]{0mm}{10mm}6 \,\, 7 &
$   \st{\bf 1}{(17)}\, \st{\bf 3}{(15)^3} \,\st{\bf 6}{(13)^5} \, (5)^6 \, (9)^3      $ \\ \hline
\rule[-2mm]{0mm}{10mm}7 \,\, 7 &
$   \st{\bf 2}{(16)^3}\, \st{\bf 5}{(14)^5} \,\st{\bf 7}{(12)^7} \, (8)^8       $ \\ \hline
\end{tabular}
\end{center}
\begin{center}
{\bf Table 2} (Continuation)
\end{center}

\newpage

\begin{center}
\begin{tabular}{||c|r|c|c||} \hline
{\it state}\rule[-2mm]{0mm}{7mm} & \multicolumn{1}{c|}{ $s/m_1^2$}  &
\multicolumn{1}{c|}{$c$--{\it series}}  & \multicolumn{1}{c||}{$U$--{\it series}} \\ \hline\hline
\rule[-2mm]{0mm}{7mm}$A_2$ &1.28558 & 0.6450605 & 0.0706975  \\ \hline
\rule[-2mm]{0mm}{7mm}$A_4$ &1.96962 &0.0256997 & 0.0066115  \\ \hline
\rule[-2mm]{0mm}{7mm}$A_1$ $A_1$ & $\geq$ 2.00000 & 0.0182735 & 0.0071135  \\ \hline
\rule[-2mm]{0mm}{7mm}$A_5$  & 2.53209 &0.0032417 & 0.0013783  \\ \hline
\rule[-2mm]{0mm}{7mm}$A_2$ $A_2$ & $\geq$ 2.57115 &0.0032549 &0.0025194   \\ \hline
\rule[-2mm]{0mm}{7mm}$A_1$ $A_3$ & $\geq$  2.87939 &0.0012782 & 0.0020630  \\ \hline
\rule[-2mm]{0mm}{7mm}$A_2$ $A_4$ & $\geq$ 3.25519 &0.0003010 & 0.0007277  \\ \hline
\rule[-2mm]{0mm}{7mm}$A_1$ $A_1$ $A_2$ & $\geq$ 3.28558 &0.0007139 & 0.001184 \\ \hline
\rule[-2mm]{0mm}{7mm}$A_7$ &3.70167 &0.0000316  & 0.0000287 \\ \hline
\rule[-2mm]{0mm}{7mm}$A_3$ $A_3$ & $\geq$ 3.75877 &0.0000700  & 0.0001173  \\ \hline
\rule[-2mm]{0mm}{7mm}$A_2$ $A_5$ & $\geq$ 3.81766 &0.0000860  & 0.0001581 \\ \hline
\multicolumn{2}{||c|}{{\it partial sum}\rule[-2mm]{0mm}{7mm} } & 
\multicolumn{1}{c|}{0.6980109}  & \multicolumn{1}{c||}{0.0914150} \\ \hline 
\multicolumn{2}{||c|}{{\it exact  value}\rule[-2mm]{0mm}{7mm} } & 
\multicolumn{1}{c|}{0.7000000}  & \multicolumn{1}{c||}{0.0942097} \\ \hline
\end{tabular}
\end{center}
\begin{center}
{\bf Table 3}
\end{center}

\newpage


\begin{center}
\begin{tabular}{||c|r||} \hline
\rule[-2mm]{0mm}{7mm}$F_2^\Theta$ & $ 0.9604936853$ \\ \hline
\rule[-2mm]{0mm}{7mm}$F_4^\Theta$ & $-0.4500141924$ \\ \hline
\rule[-2mm]{0mm}{7mm}$F_5^\Theta$ & $ 0.2641467199$ \\ \hline
\rule[-2mm]{0mm}{7mm}$F_7^\Theta$ & $-0.0556906385$ \\ \hline
\end{tabular}
\end{center}
\begin{center}
{\bf Table 4}
\end{center}

\vspace{3cm}


\begin{center}
\begin{tabular}{||c|r||} \hline
\rule[-2mm]{0mm}{7mm}$a_{11}^0$ & $6.283185307$\\ \hline
\rule[-2mm]{0mm}{7mm}$a_{13}^0$ & $30.70767637$\\ \hline
\rule[-2mm]{0mm}{7mm}$a_{22}^0$ & $15.09207695$\\ 
\rule[-2mm]{0mm}{7mm}$a_{22}^1$ & $4.707833688$\\ \hline
\rule[-2mm]{0mm}{7mm}$a_{24}^0$ & $79.32168252$\\ 
\rule[-2mm]{0mm}{7mm}$a_{24}^1$ & $16.15028004$\\ \hline
\rule[-2mm]{0mm}{7mm}$a_{33}^0$ & $295.3281130$ \\ 
\rule[-2mm]{0mm}{7mm}$a_{33}^1$ & $396.9648559$\\ 
\rule[-2mm]{0mm}{7mm}$a_{33}^2$ & $123.8295119$\\ \hline  
\rule[-2mm]{0mm}{7mm}$a_{25}^0$ & $3534.798444$\\  
\rule[-2mm]{0mm}{7mm}$a_{25}^1$ & $4062.255130$\\ 
\rule[-2mm]{0mm}{7mm}$a_{25}^2$ & $556.5589101$\\ \hline
\end{tabular}
\end{center}
\begin{center}
{\bf Table 5}
\end{center}
\newpage


\begin{center}
\begin{tabular}{||c|c|c||} \hline
{\it particle}\rule[-2mm]{0mm}{7mm} & \multicolumn{1}{c|}{ $mass/m_l$}  &
\multicolumn{1}{c||}{$Z_3$ {\it charge}}   \\ \hline\hline
\rule[-2mm]{0mm}{7mm}$A_{l}$ &1.00000 & $e^{2 \pi i/3}$  \\ \hline
\rule[-2mm]{0mm}{7mm}$A_{\lb}$ &1.00000 & $e^{-2 \pi i/3}$  \\ \hline
\rule[-2mm]{0mm}{7mm}$A_{L}$ &1.41421 & $1$  \\ \hline
\rule[-2mm]{0mm}{7mm}$A_{h}$ &1.93185 & $e^{2 \pi i/3}$  \\ \hline
\rule[-2mm]{0mm}{7mm}$A_{\hb}$ &1.93185 & $e^{-2 \pi i/3}$  \\ \hline
\rule[-2mm]{0mm}{7mm}$A_{H}$ &  2.73205 & $1$  \\ \hline
\end{tabular}
\end{center}
\begin{center}
{\bf Table 6}
\end{center}
\newpage

\begin{center}
\begin{tabular}{|c|c|}\hline
$a$ \,\, $b$ &
$S_{ab}$ \\ \hline \hline
\rule[-2mm]{0mm}{10mm}$l \,\,\,\,\, l $&
$   \st{ \lb}{[8]}  \, [6] \, \st{ \hb}{[2]} $\\ \hline
\rule[-2mm]{0mm}{10mm}$\lb \,\,\,\,\, \lb $&
$   \st{ l}{[8]}  \, [6] \, \st{ h}{[2]} $\\ \hline
\rule[-2mm]{0mm}{10mm}$l \,\,\,\,\, \lb $&
$  - [10] \, \st{L}{[6]}  \, [4]  $\\ \hline
\rule[-2mm]{0mm}{10mm}$l \,\,\,\,\, L $&
$    \st{ l}{[9]} \,[7] \,  \st{ h}{[5]} \, [3]  $\\ \hline
\rule[-2mm]{0mm}{10mm}$\lb \,\,\,\,\, L $&
$    \st{ \lb}{[9]} \,[7] \,  \st{ \hb}{[5]} \, [3]  $\\ \hline
\rule[-2mm]{0mm}{10mm}$l \,\,\,\,\, h $&
$  [9] \,  \st{ \hb}{[7]} \,[5]^2 \, [3]\, \st{ \lb}{[11]}   $\\ \hline
\rule[-2mm]{0mm}{10mm}$\lb \,\,\,\,\, \hb $&
$  [9] \,  \st{ h}{[7]} \,[5]^2 \, [3]\, \st{ l}{[11]}   $\\ \hline
\rule[-2mm]{0mm}{10mm}$l \,\,\,\,\, \hb $&
$    \st{ L}{[9]} \,[7]^2 \, [5]\, \st{ H}{[3]} \, [1]  $\\ \hline
\rule[-2mm]{0mm}{10mm}$\lb \,\,\,\,\, h $&
$    \st{ L}{[9]} \,[7]^2 \, [5]\, \st{ H}{[3]} \, [1]  $\\ \hline
\rule[-2mm]{0mm}{10mm}$l \,\,\,\,\, H $&
$    \st{ h}{[10]} \,[8]^2 \,[6]^2\,[4]^2\, [2]  $\\ \hline
\rule[-2mm]{0mm}{10mm}$\lb \,\,\,\,\, H $&
$    \st{ \hb}{[10]} \,[8]^2 \,[6]^2\,[4]^2\, [2]  $\\ \hline
\rule[-2mm]{0mm}{10mm}$L \,\,\,\,\, L $&
$   - [10] \, \st{ L}{[8]} \,[6]^2 \,[4] \,\st{ H}{[2]}    $\\ \hline
\rule[-2mm]{0mm}{10mm}$L \,\,\,\,\, h $&
$    \st{ l}{[10]} \,[8]^2 \,[6]^2\,[4]^2\, [2]  $\\ \hline
\rule[-2mm]{0mm}{10mm}$L \,\,\,\,\, \hb $&
$    \st{ \lb}{[10]} \,[8]^2 \,[6]^2\,[4]^2\, [2]  $\\ \hline
\rule[-2mm]{0mm}{10mm}$L \,\,\,\,\, H $&
$    \st{ L}{[11]} \,[9]^2 \,\st{ H}{[7]^3}  \,[5]^3\, [3]^2 \, [1]  $\\ \hline
\rule[-2mm]{0mm}{10mm}$h \,\,\,\,\, h $&
$    \st{ \lb}{[10]} \,\st{ \hb}{[8]^3}  \,[6]^3\,[4]^2\, [2]^2  $\\ \hline
\rule[-2mm]{0mm}{10mm}$\hb \,\,\,\,\, \hb $&
$    \st{ l}{[10]} \,\st{ h}{[8]^3}  \,[6]^3\,[4]^2\, [2]^2  $\\ \hline
\rule[-2mm]{0mm}{10mm}$h \,\,\,\,\, \hb $&
$  -  [10]^2 \,[8]^2\,\st{ H}{[6]^3}  \,[4]^3\, [2]  $\\ \hline
\rule[-2mm]{0mm}{10mm}$h \,\,\,\,\, H $&
$    \st{ l}{[11]} \,\st{ h}{[9]^3}  \,[7]^4\,[5]^4\, [3]^3 \, [1]  $\\ \hline
\rule[-2mm]{0mm}{10mm}$\hb \,\,\,\,\, H $&
$    \st{ \lb}{[11]} \,\st{ \hb}{[9]^3}  \,[7]^4\,[5]^4\, [3]^3 \, [1]  $\\ \hline
\rule[-2mm]{0mm}{10mm}$H \,\,\,\,\, H $&
$  -  \st{ L}{[10]^3}  \,\st{ H}{[8]^5}  \,[6]^6\, [4]^5 \, [2]^3  $\\ \hline
\end{tabular}
\end{center}
\begin{center}
{\bf Table 7}
\end{center}
\newpage


\begin{center}
\begin{tabular}{||c|r|l|l||} \hline
{\it state}\rule[-2mm]{0mm}{7mm} & \multicolumn{1}{c|}{ $s/m_1^2$}  &
\multicolumn{1}{c|}{$c$--series}  & \multicolumn{1}{c||}{$u$--series} \\ \hline\hline
\rule[-2mm]{0mm}{7mm}$A_L$ &1.41421 & 0.7596531 & 0.0705265  \\ \hline
\rule[-2mm]{0mm}{7mm}$A_l$ $A_{\lb}$ & $\geq$ 2.00000 &0.0844238 & 0.0229507  \\ \hline
\rule[-2mm]{0mm}{7mm}$A_H$ &  2.73205 & 0.0029236 & 0.001013  \\ \hline
\rule[-2mm]{0mm}{7mm}$A_L$ $A_L$  & $\geq$ 2.82843 &0.0024419 & 0.0019380  \\ \hline
\rule[-2mm]{0mm}{7mm}$A_l$ $A_{\hb}$   & $\geq$ 2.93185 & 0.0023884  &  0.0016745    \\ \hline
\rule[-2mm]{0mm}{7mm}$A_{\lb}$ $A_h$   & $\geq$ 2.93185 & 0.0023884  &  0.0016745   \\ \hline
\rule[-2mm]{0mm}{7mm}$A_l$ $A_l$  $A_l$   & $\geq$  3.00000 &  0.0004215 &  0.0004925  \\ 
\hline 
\rule[-2mm]{0mm}{7mm} $A_{\lb}$ $A_{\lb}$  $A_{\lb}$  & $\geq$  3.00000 &  0.0004215 &  0.0004925  \\ 
\hline 
\rule[-2mm]{0mm}{7mm}$A_l$ $A_{\lb}$ $A_L$ & $\geq$ 3.41421 &0.00159 & 0.000251  \\ \hline
\rule[-2mm]{0mm}{7mm}$A_h$  $A_{\hb}$ & $\geq$ 3.86370 &0.0000504 & 0.0001476 \\ \hline
\rule[-2mm]{0mm}{7mm}$A_l$ $A_l$ $A_h$  &$\geq$ 3.93185 &0.000089  &  0.0002015 \\ 
\hline
\rule[-2mm]{0mm}{7mm} $A_{\lb}$ $A_{\lb}$ $A_{\hb}$ &$\geq$ 3.93185 &0.000089  &  0.0002015 \\ 
\hline
\rule[-2mm]{0mm}{7mm}$A_l$ $A_{\lb}$  $A_l$ $A_{\lb}$ & $\geq$ 4.00000 &0.0000959  & 0.000381  \\ \hline
\multicolumn{2}{||c|}{{\it partial sum}\rule[-2mm]{0mm}{7mm} } & 
\multicolumn{1}{c|}{0.8569765}  & \multicolumn{1}{c||}{0.1019449} \\ \hline 
\multicolumn{2}{||c|}{{\it exact  value}\rule[-2mm]{0mm}{7mm} } & 
\multicolumn{1}{c|}{0.8571429}  & \multicolumn{1}{c||}{0.1056624} \\ \hline
\end{tabular}
\end{center}
\begin{center}
{\bf Table 8}
\end{center}
\newpage


\begin{center}
\begin{tabular}{||c|r||} \hline
\rule[-2mm]{0mm}{7mm}$F_L^\Theta$ & $1.261353947 $ \\ \hline
\rule[-2mm]{0mm}{7mm}$F_H^\Theta$ & $0.292037405$ \\ \hline
\end{tabular}
\end{center}
\begin{center}
{\bf Table 9}
\end{center}

\vspace{3cm}


\begin{center}
\begin{tabular}{||c|r||} \hline
\rule[-2mm]{0mm}{7mm}$a_{l\,\lb}^0$ & $6.283185307$\\ \hline
\rule[-2mm]{0mm}{7mm}$a_{L\,L}^0$   & $21.76559237$\\ 
\rule[-2mm]{0mm}{7mm}$a_{L\,L}^1$   & $9.199221756$\\ \hline
\rule[-2mm]{0mm}{7mm}$a_{l\,\hb}^0$ & $25.22648264$\\ \hline
\rule[-2mm]{0mm}{7mm}$a_{h\,\hb}^0$ & $414.1182423$\\  
\rule[-2mm]{0mm}{7mm}$a_{h\,\hb}^1$ & $565.6960386$\\ 
\rule[-2mm]{0mm}{7mm}$a_{h\,\hb}^2$ & $175.0269632$\\ \hline
\end{tabular}
\end{center}
\begin{center}
{\bf Table 10}
\end{center}
\newpage

\vspace{25mm}

{\bf Figure Captions}

\vspace{1cm}

\begin{description}
\item [Figure 1] Diagrammatic interpretation of the process responsible for a single--pole
in a Form Factor. 
\item [Figure 2] Diagrammatic interpretation of the process responsible for a double--pole
in a Form Factor. 
\item [Figure 3] Diagrammatic interpretation of the process responsible for a triple--pole
in a Form Factor (here $\varphi= u_{ab}^{f}$). 
\item [Figure 4]   Dynkin diagram of $E_7$ and assignment of the
 masses to the corresponding dots. 
\item [Figure 5] Plot of the  correlation function 
$\langle\,\Theta(x) \Theta(0)\,\rangle /m_1^4$ 
versus the scaling variable $m_1\,|x|$ 
in the TIM. The spectral series (\ref{formexp}) includes the  FF contributions
relative to the multiparticle states  in Table  3. 
\item [Figure 6]  Dynkin diagram of $E_6$ and assignment of the
 masses to the corresponding dots.
\item [Figure 7] Plot of the  correlation function 
$\langle\,\Theta(x) \Theta(0)\,\rangle /m_l^4$ 
versus the scaling variable $m_l\,|x|$ 
in the TPM. The spectral series (\ref{formexp}) includes the  FF contributions
relative to the multiparticle states  in Table  8.
\end{description}

\begin{figure}[h]
\vspace{5cm}
\centerline{\psfig{figure=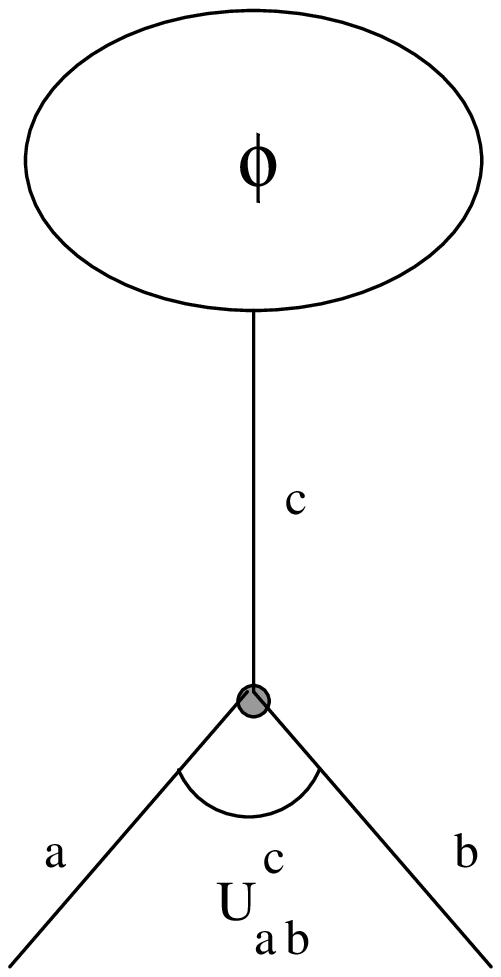}}
\vspace{3cm}
\begin{center}
{\bf Figure 1}
\end{center}
\end{figure}

\begin{figure}[h]
\vspace{5cm}
\centerline{\hspace{2cm}\psfig{figure=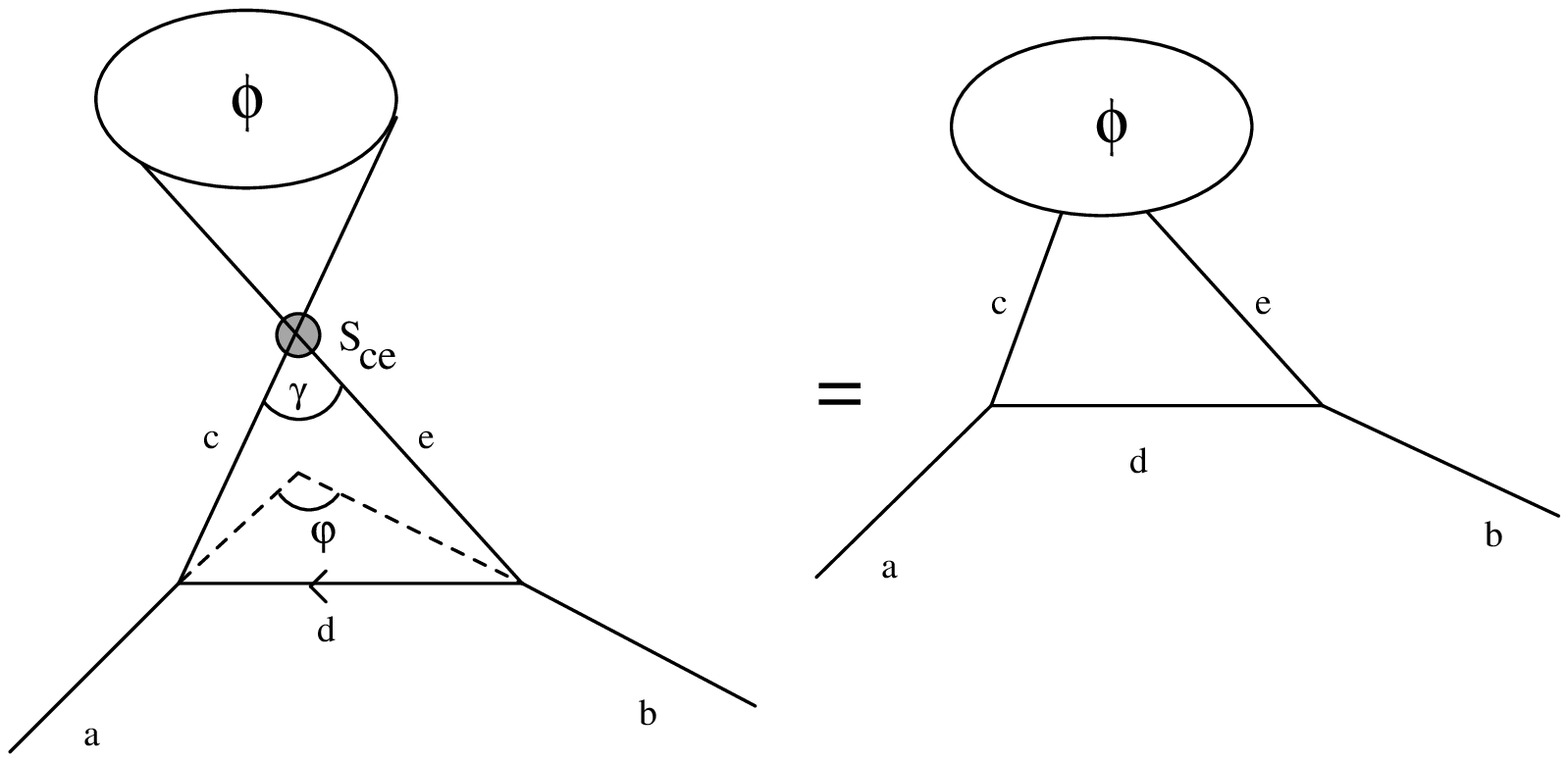}}
\vspace{3cm}
\begin{center}
{\bf Figure 2}
\end{center}
\end{figure}

\begin{figure}[h]
\vspace{5cm}
\centerline{\hspace{2cm}
\psfig{figure=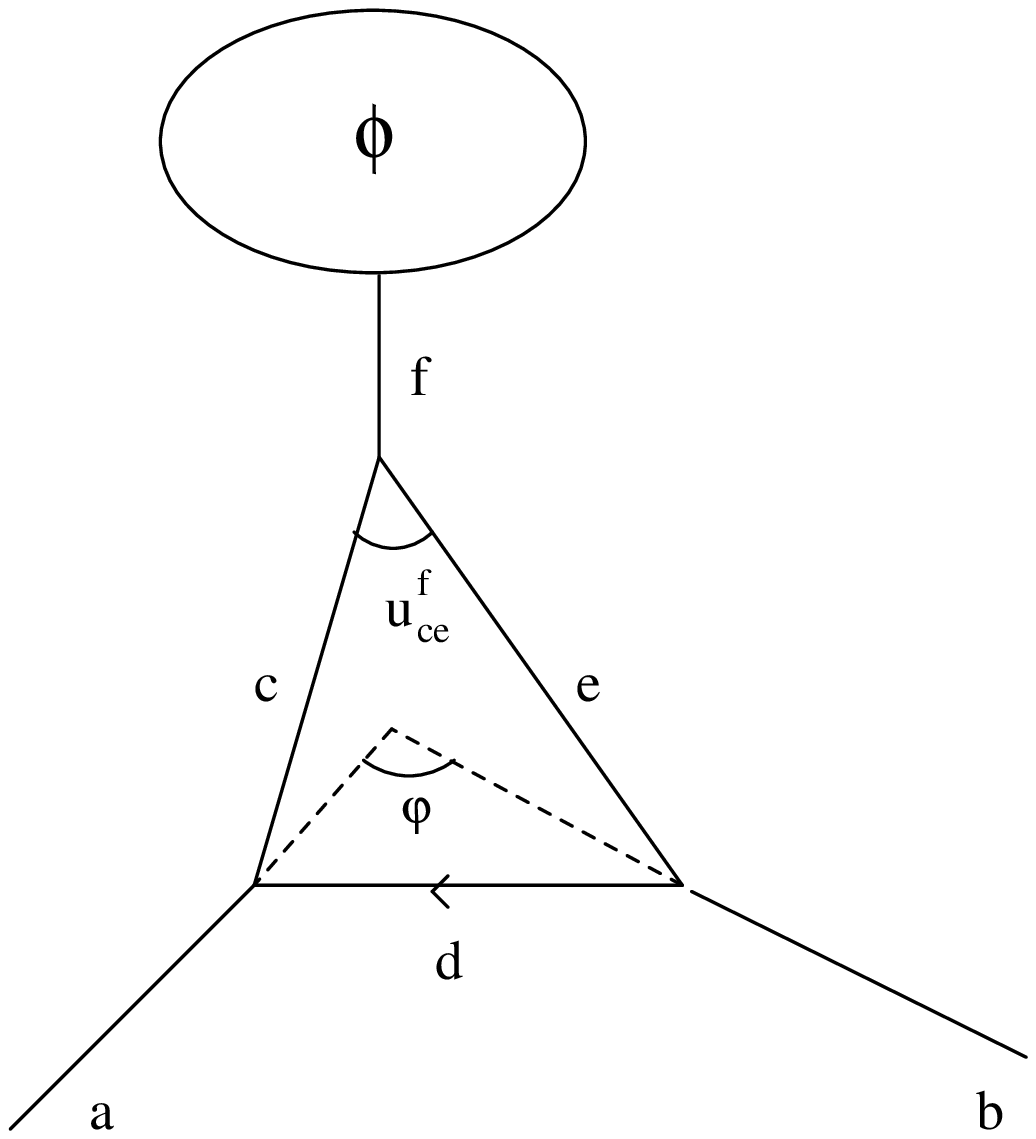}}
\vspace{3cm}
\begin{center}
{\bf Figure 3}
\end{center}
\end{figure}

\begin{figure}
\begin{center}
\begin{picture}(240,130)
\thicklines
\put(0,30){\line(1,0){240}}
\put(0,30){\line(0,1){100}}
\put(240,30){\line(0,1){100}}
\put(0,130){\line(1,0){240}}
\put(80,70){\line(1,0){100}}
\put(120,70){\line(0,1){20}}
\put(80,70){\circle*{3}} 
\put(100,70){\circle*{3}} 
\put(120,70){\circle*{3}} 
\put(140,70){\circle*{3}} 
\put(160,70){\circle*{3}} 
\put(180,70){\circle*{3}} 
\put(120,90){\circle*{3}} 
\put(80,60){\makebox(0,0){$m_2$}}
\put(100,60){\makebox(0,0){$m_5$}}
\put(120,60){\makebox(0,0){$m_7$}}
\put(140,60){\makebox(0,0){$m_6$}}
\put(160,60){\makebox(0,0){$m_4$}}
\put(180,60){\makebox(0,0){$m_1$}}
\put(110,90){\makebox(0,0){$m_3$}}
\end{picture}
\end{center}
\begin{center}
{\bf Figure 4}
\end{center}
\end{figure}

\begin{figure}
\centerline{\psfig{figure=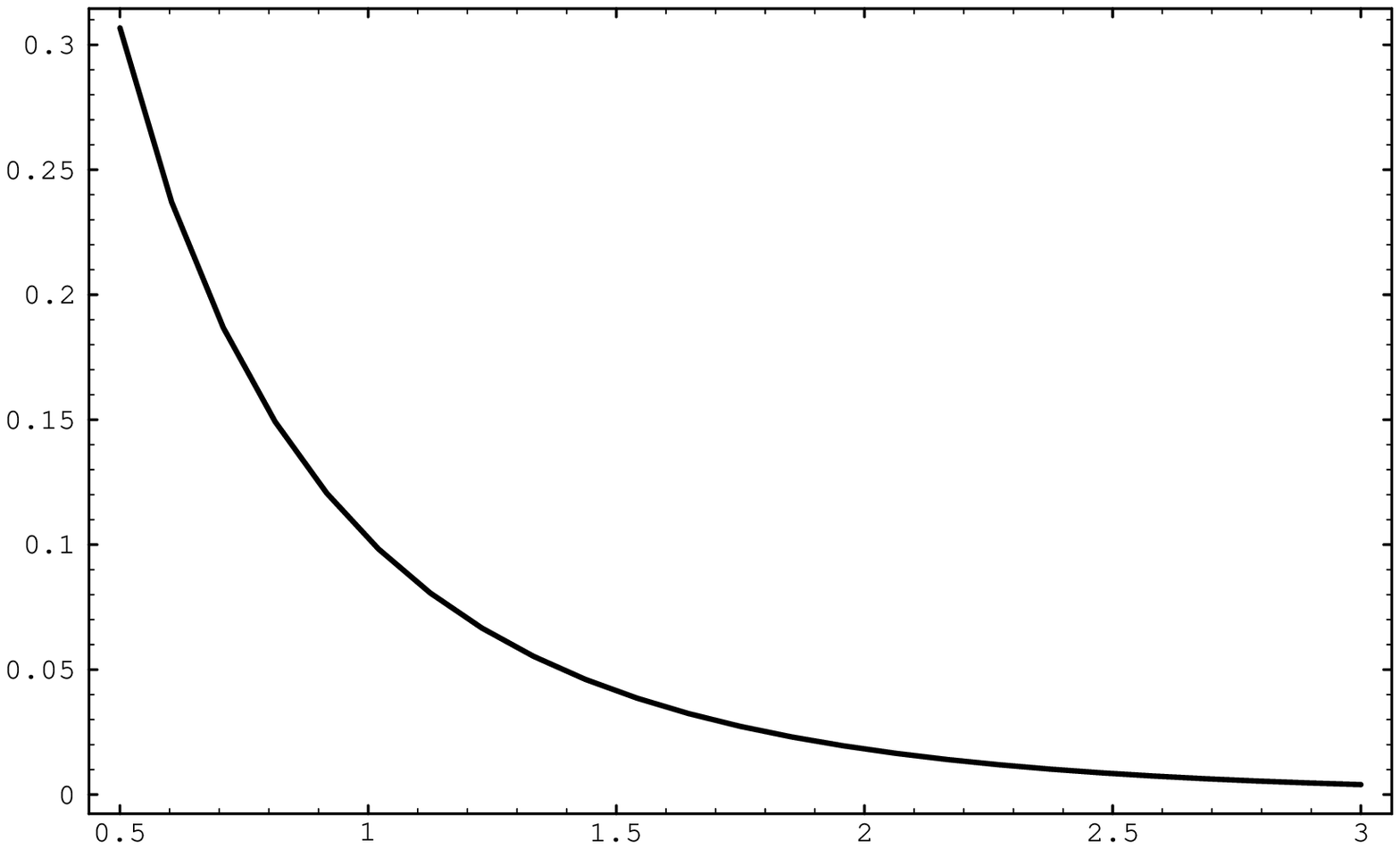}}
\vspace{-5cm}
\begin{center}
{\bf Figure 5}
\end{center}
\end{figure}

\begin{figure}
\begin{center}
\begin{picture}(240,130)
\thicklines
\put(0,30){\line(1,0){240}}
\put(0,30){\line(0,1){100}}
\put(240,30){\line(0,1){100}}
\put(0,130){\line(1,0){240}}
\put(80,70){\line(1,0){80}}
\put(120,70){\line(0,1){20}}
\put(80,70){\circle*{3}} 
\put(100,70){\circle*{3}} 
\put(120,70){\circle*{3}} 
\put(140,70){\circle*{3}} 
\put(160,70){\circle*{3}} 
\put(120,90){\circle*{3}} 
\put(80,60){\makebox(0,0){$m_l$}}
\put(100,60){\makebox(0,0){$m_h$}}
\put(120,60){\makebox(0,0){$m_H$}}
\put(140,60){\makebox(0,0){$m_{\ov{h}}$}}
\put(160,60){\makebox(0,0){$m_{\ov{b}}$}}
\put(110,90){\makebox(0,0){$m_L$}}
\end{picture}
\end{center}
\begin{center}
{\bf Figure 6}
\end{center}
\end{figure}

\begin{figure}
\centerline{\psfig{figure=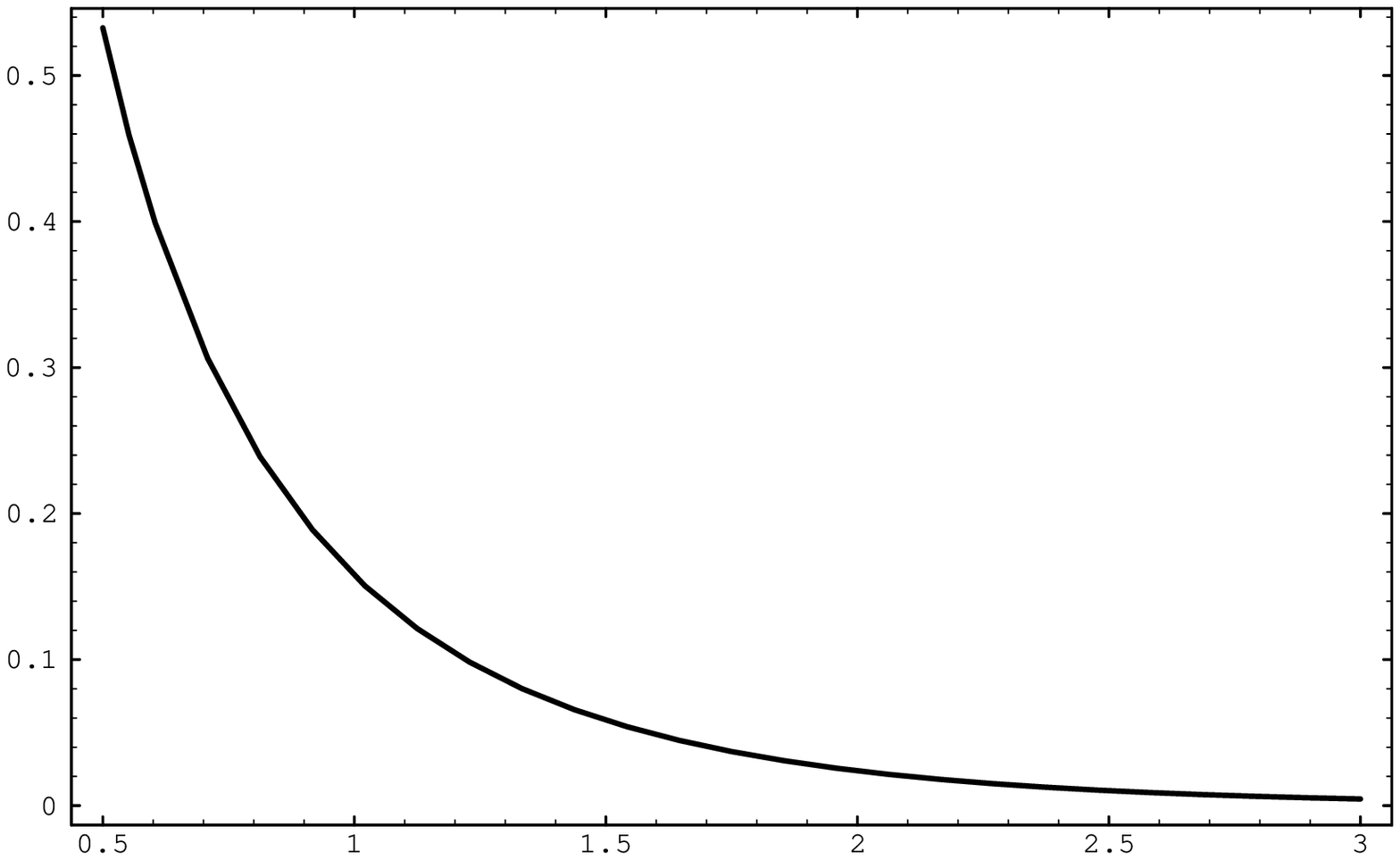}}
\vspace{-5cm}
\begin{center}
{\bf Figure 7}
\end{center}
\end{figure}

\end{document}